%% file: prd_ajets_v9.tex
\documentstyle[twoside,12pt,epsf,aps]{revtex}
 \relax
 \input{prd_definitions}
  \begin{document}
  \bibliographystyle{unsrt}
  \title{Heavy flavor properties of jets produced in $p\bar{p}$
         interactions at $\sqrt{s}$= 1.8 TeV}
  \maketitle
  \input{auth_luc.tex}

  \begin{abstract}
   We present a detailed examination of the heavy flavor properties of jets
   produced at the Fermilab Tevatron collider. The data set, collected with 
   the Collider Detector at Fermilab, consists of events with two or more 
   jets with transverse energy $E_T \geq 15\; \gev$ and pseudo-rapidity 
   $|\eta| \leq 1.5$. The heavy flavor content of the data set is enriched by
   requiring that at least one of the jets (lepton-jet) contains a lepton with
   transverse momentum larger than $8 \; \gevc$. Jets containing hadrons with
   heavy flavor are selected via the identification of secondary vertices.
   The parton-level cross sections predicted by the {\sc herwig} Monte Carlo
   generator program are tuned within theoretical and experimental 
   uncertainties to reproduce the secondary-vertex rates in the data.
   The tuned simulation provides new information on the origin of the 
   discrepancy between the $b\bar{b}$ cross section measurements at the 
   Tevatron and the next-to-leading order QCD prediction. We also compare the
   rate of away-jets (jets recoiling against the lepton-jet) containing a soft
   lepton ($p_T \geq 2\; \gevc$) in the data to that in the tuned simulation.
   We find that this rate is larger than what is expected for the conventional
   production and semileptonic decay of pairs of hadrons with heavy flavor.\\
   PACS number(s): 13.85.Qk, 13.20.He, 13.20.Fc
 \end{abstract} 
  \section {Introduction} \label{sec:ss-intro} 
   This paper presents a study of semileptonic decays in jets containing
   heavy flavor and is motivated by several anomalies that have been 
   previously reported. CDF has found the rate of jets with both a secondary
   vertex and a soft lepton (superjets) to be larger than expected in the
   $\W+$ 2,3 jet sample. The kinematical properties of the events with a
   superjet are difficult to reconcile with the standard model (SM) 
   expectation~\cite{suj}.

   The discrepancy between the single bottom quark cross section measurements
   at the Tevatron and the next-to-leading order (NLO) QCD 
   prediction~\cite{nde} can be explained either in terms of new 
   physics~\cite{berger} or by the lack of robustness of the NLO 
   prediction~\cite{braaten}. However, at the Tevatron, there are two 
   additional discrepancies between the measured and predicted value of the
   $b\bar{b}$ cross section that are more difficult to accommodate within
   the theoretical uncertainty. In Ref.~\cite{derwent}, the correlated 
   $\mu+\bar{b}$-jet cross section is measured to be 1.5 times larger 
   than $\sigma_{b\bar{b}} \times BR$, where $BR$ is the average
   semileptonic branching ratio of $b$-hadrons produced at the Tevatron 
   and $\sigma_{b\bar{b}}$ is the NLO prediction of the cross section for 
   producing pairs of $b$ and $\bar{b}$ quarks. A further discrepancy is
   found by both CDF and D${\not \! {\rm O}}$ experiments~\cite{yuntae,abbot}
   when comparing the cross section for producing dimuons from $b$-hadron
   semileptonic decays to $\sigma_{b\bar{b}} \times {BR}^2$.
   The value of $\sigma_{b\bar{b}} \times {BR}^2$ is found to be approximately
   2.2 times larger than the NLO prediction~\footnote
   {In both measurements, $\sigma_{b\bar{b}}$ is the cross section for
    producing two central bottom quarks, both with transverse momentum
    approximately larger than $10\; \gevc$. In this case, the LO and NLO 
    predictions are equal within a few percents, and the NLO prediction 
    changes by no more than 15\% when changing the renormalization and
    factorization scales by a factor of two~\cite{mnr}.}.
   There are possible conventional explanations presented in the literature 
   for these anomalies~\cite{frix,jung}.

   However, all these discrepancies could also be mitigated by postulating the
   existence of a light strong-interacting object with a 100\% semileptonic
   branching ratio. Since there are no limits to the existence of a
   charge$-1/3$ scalar quark with mass smaller than 
   $7.4\; \gevcc$~\cite{mssm-can,nappi,carena}, the supersymmetric partner of
   the bottom quark is a potential candidate. This paper presents an analysis
   of multi-jet data intended to search for evidence either supporting or 
   disfavoring this hypothesis.

   The strategy of this search is outlined in Sec.~\ref{sec:strat}. 
   Section~\ref{sec:ss-det} describes the detector systems relevant to this
   analysis, while the sample selection and the tagging algorithms (SECVTX 
   and JPB) used to select heavy flavors are discussed in
   Sec.~\ref{sec:ss-sample}. Section~\ref{sec:ss-comp} describes the data
   sample composition and the heavy flavor simulation. The data set consists
   of events with two or more jets with transverse energy $E_T \geq 15\; \gev$
   and contained in the silicon microvertex detector (SVX) acceptance. 
   The sample is enriched in heavy flavor by requiring that at least one of 
   the jets contains a lepton with $p_T \geq 8\; \gevc$. We use measured
   rates of SECVTX and JPB tags to determine the bottom and charmed content
   of the data; we then tune the simulation to match the heavy-flavor content
   of the data. The evaluation of the number of SECVTX and JPB tags due to
   heavy flavor in the data and the simulation is described in
   Sec.~\ref{sec:ss-datatag} and~\ref{sec:ss-simtag}, respectively. 
   The tuning of the heavy flavor production cross sections in the simulation
   is described in Sec.~\ref{sec:ss-norm}. In Sec.~\ref{sec:ss-rateslt}
   we measure the yields of jets containing soft leptons 
   ($p_T \geq 2 \; \gevc$), and compare them to the prediction of the tuned
   simulation. Section~\ref{sec:ss-syst} contains cross-checks and a
   discussion of the systematic uncertainties. Our conclusions are
   presented in Sec.~\ref{sec:concl}.
  \section{Probing the production of light scalar quarks with a large 
           semileptonic branching ratio}\label{sec:strat}
   In previous publications~\cite{suj,cdf-tsig} we have compared the $b$-
   and $c$-quark content of several samples of generic-jet data to the
   QCD prediction of the standard model  using the {\sc herwig} generator
   program~\cite{herwig}. We identify (tag) jets produced by heavy quarks
   using the CDF silicon micro-vertex detector (SVX) to locate secondary
   vertices produced by the decay of $b$ and $c$ hadrons inside a jet. 
   These vertices (SECVTX tags) are separated from the primary event 
   vertex as a result of the long $b$ and $c$ lifetime. We also use track
   impact parameters to select jets with a small probability of originating
   from the primary vertex of the event (JPB tags)~\cite{jpb}.

   In Ref.~\cite{cdf-tsig} we have compared rates of SECVTX and JPB tags
   in generic-jet data and their simulation first to calibrate the
   efficiency of the tagging algorithms in the simulation, and then to
   tune the heavy flavor cross sections evaluated with the {\sc herwig}
   parton shower Monte Carlo. In the simulation, jets with heavy flavor
   are produced by heavy quarks in the initial or final state of the hard
   scattering (flavor excitation and direct production, respectively) or
   from gluons branching into $b\bar{b}$ or $c\bar{c}$ pairs (gluon
   splitting). The fraction of generic-jet data containing $b\bar{b}$ or
   $c\bar{c}$ pairs calculated by {\sc herwig} models correctly the
   observed rate of tags after minor adjustments within the theoretical
   and experimental uncertainties. In Refs.~\cite{suj,cdf-tsig}, we have
   extended this comparison to $\W+$ jet events. We find again good
   agreement between the observed rates of SECVTX and JPB tags and the SM
   prediction, which includes single and pair production of top quarks.

   We also identify heavy flavors by searching jets for leptons ($e$ or
   $\mu$) produced in the decay of $b$ and $c$ hadrons~\cite{suj,cdf-tsig};
   we refer to these as soft lepton tags (SLT). As shown in
   Refs.~\cite{suj,cdf-tsig}, rates of SLT tags in generic-jet data and 
   in $\W +$ jet events are generally well modeled by the simulation.
   An exception is the rate of SECVTX+SLT tags in the same jet (called
   supertags in Ref.~\cite{suj}) that, in $\W+$ 2,3 jet events, is larger
   than in the simulation, whereas, in generic-jet samples, is slightly
   overpredicted by the same simulation.

   This analysis uses two data samples, referred to as the signal or
   inclusive lepton sample and the control or generic-jet sample. The signal
   sample consists of events with two or more jets that have been acquired
   with the trigger request that events contain a lepton with 
   $p_T \geq 8 \; \gevc$. The request of a jet containing a lepton
   (lepton-jet) enriches the heavy flavor content of the sample with respect
   to generic jets. The control or generic-jet sample is the same sample
   studied in Refs.~\cite{suj,cdf-tsig}, and consists of events with one 
   or more jets acquired with three trigger thresholds of $20$, $50$ and 
   $100\; \gev$, respectively.

   In the signal sample, we study  jets recoiling against the lepton-jet
   (away-jets) and we perform three measurements: we count the number of 
   away-jets that contain a lepton (SLT tag); that contain an SLT tag and a 
   SECVTX tag; that contain an SLT tag and a JPB tag. The latter two are
   referred to as supertags. We compare the three measurements to a Monte
   Carlo simulation which is tuned and normalized to the data by equalizing
   numbers of SECVTX and JPB tags. The normalization and tuning procedure
   serves two purposes: it removes the dependence on the efficiency for 
   finding the trigger lepton and ensures that the simulation reproduces 
   the heavy-flavor content of the data, respectively. To calibrate the 
   efficiency for finding SLT tags or supertags in the simulation, we use
   rates of SLT tags and supertags in generic-jet data (control sample). 
   In Ref.~\cite{suj}, we have compared these measurements to a Monte Carlo
   simulation which was also tuned and normalized to generic-jet data by
   equalizing numbers of SECVTX and JPB tags. These three comparisons are
   used to verify the simulated efficiency for finding SLT tags, and to
   empirically calibrate the efficiency for finding supertags in the 
   simulation.

   This analysis strategy is motivated by the following argument. If 
   low-mass bottom squarks existed, they would be produced copiously at
   the Tevatron. The NLO calculation of the process $p \bar{p}
   \rightarrow \tilde{b} \tilde{b}^{*}$, implemented in the {\sc prospino}
   Monte Carlo generator~\cite{prosp}, predicts a cross section which is
   $\simeq 15$\% of the NLO prediction for the production cross section
   of quarks with the same mass~\cite{mnr}. In Ref.~\cite{cdf-tsig}, we have
   tuned, within the theoretical and experimental uncertainties, the heavy 
   flavor production cross sections calculated by {\sc herwig} to reproduce
   the rates of SECVTX and JPB tags observed in generic-jet data. However,
   if the squark lifetime is similar to that of conventional heavy flavors,
   we have unfortunately tuned the parton-level cross section evaluated
   by {\sc herwig} (or the number of simulated SECVTX and JPB tags predicted
   by the simulation) to explain in terms of conventional processes the
   squark production. However, if bottom squarks have a 100\% semileptonic
   branching ratio, it is still possible to identify their presence by 
   comparing the observed number of jets containing a lepton to that
   expected from $b$ and $c$ decays.

   We illustrate the procedure used in this paper with a numeric example
   detailed in Table~\ref{tab:tab_strat}. The first column is what there 
   would be in the data in the presence of $\tilde{b}$ quarks with 100\%
   semileptonic $BR$~\footnote{
   The cross sections are predicted using the {\sc mnr}~\cite{mnr} and
   {\sc prospino}~\cite{prosp} Monte Carlo generators, the MRS(G) set of
   structure functions~\cite{mrsg}, and the renormalization and
   factorization scales $\mu_0^{2}=p_T^{2}+m_{\tilde{b}}^{2}$. We use
   $m_b$ = 4.75 $\gevcc$,  $m_c$ = 1.5 $\gevcc$, and 
   $m_{\tilde{b}}$ = 3.6 $\gevcc$. The cross section are integrated over
   final-state partons with $p_T \geq 18 \; \gevc$; this threshold is used 
   to mimic the generic-jet data. Bottom quarks have a 37\% semileptonic
   branching ratio, $BR$, due to $b \rightarrow l $ and 
   $b \rightarrow c \rightarrow l$ decays, whereas $BR= 21$\% for
   $c$ quarks~\cite{hf-sem}.}.
   The cross sections in the first column of row A represent approximately
   the different heavy flavor contributions to the generic-jet sample.
   The second column  is what one would predict after having tuned a 
   simulation, in which only $b$ and $c$ quarks are present, to reproduce
   the number of SECVTX and JPB  tags observed in the sample corresponding
   to the first column of row A, in the assumption that $b$ and $\tilde{b}$
   quarks have the same lifetime. In row B, we model the request that a jet
   contains a lepton by multiplying the heavy flavor cross sections by the
   respective semileptonic branching ratios $BR$. A 20\% excess is observed.
   In row C, we mimic the case in which two jets contain a lepton, and the
   same analysis leads to an excess of a factor of two. Since a discrepancy
   that depends on the number of leptons could be due to a wrong simulation
   of the lepton-identification efficiency, row D presents the stratagem
   of tuning again the conventional heavy flavor cross sections for 
   producing events with one lepton (second column in row B) to model the
   cross section contributing to events with one lepton (first column in
   row B)~\footnote{
   This technique also allows us to use the inclusive lepton sample that
   corresponds to a much larger integrated luminosity than that of 
   generic-jet data.}.
   Next, row E shows the result of requiring an additional lepton in 
   sample D: the excess is a factor of 1.5. If one chooses, as we did in
   previous studies, to use sample B to empirically correct the simulated
   efficiency for identifying a lepton, sample E will show a 30\% excess.
  \input{tab_strat.tex}

  \section{The CDF detector} \label{sec:ss-det}
   The events used for this analysis have been collected with the CDF 
   detector during the $1993-1995$ run of the Tevatron collider at Fermilab.
   The CDF detector is described in detail in Ref.~\cite{cdf-det}. We review
   the detector components most relevant to this analysis. Inside the 1.4 T 
   solenoid the silicon microvertex detector (SVX)~\cite{svx-det}, a vertex 
   drift chamber (VTX), and the central tracking chamber (CTC) provide the
   tracking and momentum information for charged particles. The CTC is a
   cylindrical drift chamber containing 84 measurement layers. It covers
   the pseudo-rapidity interval $|\eta|\leq 1.1$, where 
   $\eta=-\ln [\tan(\theta/2)]$. In CDF, $\theta$ is the polar angle measured
   from the proton direction, $\phi$ is the azimuthal angle, and $r$ is is 
   the radius from the beam axis ($z$-axis). The SVX consists of four layers
   of silicon micro-strip detectors, located at radii between 2.9 and 7.9 cm
   from the beam line, and provides spatial measurements in the $r-\phi$ 
   plane with a resolution of 13 $\mu$m.

   Electromagnetic (CEM) and hadronic (CHA) calorimeters with projective 
   tower geometry are located outside the solenoid and cover the 
   pseudo-rapidity region $|\eta|\leq 1.1$, with a segmentation of 
   $\Delta \phi =15^{\deg}$ and $\Delta \eta=0.11$. A layer of proportional
   chambers (CES) is embedded near shower maximum in the CEM and provides a
   more precise measurement of the electromagnetic shower position. Two muon
   subsystems in the central rapidity region ($|\eta|\leq 0.6$) are used for
   muon identification: the central muon chambers (CMU), located behind the
   CHA calorimeter, and the central upgrade muon chambers (CMP), located 
   behind an additional 60 cm of steel. The central muon extension (CMX)
   covers approximately 71\% of the solid angle for $0.6 \leq|\eta|\leq 1.0$
   and, in this analysis, is used only to increase the soft muon acceptance.

   CDF uses a three-level trigger system. At the first two levels, decisions
   are made with dedicated hardware. The information available at this stage
   includes energy deposited in the CEM and CHA calorimeters, high-$p_T$ 
   tracks found in the CTC by a fast track processor (CFT), and track 
   segments found in the muon subsystems. The data used in this study were
   collected using the electron and muon low-$p_T$ triggers. The first two 
   levels of these triggers require a track with $p_T \geq 7.5\; \gevc$ 
   found by the CFT. In the case of the electron trigger, the CFT track must
   be matched to a CEM cluster with transverse energy $E_T \geq 8\; \gev$. 
   In the case of the muon trigger, the CFT track must be matched to a
   reconstructed track-segment in both sets of central muon detectors
   (CMU and CMP).

   At the third level of the trigger, the event selection is based on a 
   version of the off-line reconstruction programs optimized for speed. The
   lepton selection criteria used by the third level trigger are similar to
   those described in the next section.
  \section{Data sample selection and heavy flavor tagging}
  \label{sec:ss-sample}
   Central electrons and muons that passed the trigger prerequisite are 
   identified with the same criteria used to select the $\W+$ jet sample
   described in Refs.~\cite{suj,cdf-tsig}.
 
   Electron candidates are identified using information from both
   calorimeter and tracking detectors. We require the following:
   (1) the ratio of hadronic to electromagnetic energy of the cluster,
   $E_{had}/E_{em}\leq 0.05$; (2) the ratio of cluster energy to track
   momentum, $E/p \leq 1.5$; (3) a comparison of the lateral shower profile
   in the calorimeter cluster with that of test-beam electrons, 
   $L_{shr} \leq 0.2$; (4) the distance between the extrapolated 
   track-position and the CES measurement in the $r-\phi$ and $z$ views, 
   $\Delta x \leq 1.5\; {\rm cm}$ and $\Delta z \leq 3.0\; {\rm cm}$, 
   respectively; (5) a $\chi^{2}$ comparison of the CES shower profile with 
   those of test-beam electrons, $\chi^{2}_{strip} \leq 20$; (6) the distance
   between the interaction vertex and the reconstructed track in the
   $z$-direction, $z$-vertex match $\leq$ 5 cm. Fiducial cuts on the
   electromagnetic shower position, as measured in the CES, are applied to
   ensure that the electron candidate is away from the calorimeter boundaries
   and the energy is well measured. Electrons from photon conversions are
   removed using an algorithm based on track information~\cite{cdf-tsig}.

   Muons are identified by requiring a match between a CTC track and track
   segments in both the CMU and CMP muon chambers. The following variables
   are used to separate muons from hadrons interacting in the calorimeter
   and cosmic rays: (1) an energy deposition in the electromagnetic and
   hadronic calorimeters characteristic of minimum ionizing particles,
   $E_{em} \leq 2\; \gev$ and $E_{had}\leq 6\; \gev$, respectively; 
   (2) $E_{em} + E_{had} \geq 0.1\; \gev$; 
   (3) the distance of closest approach of the reconstructed track to the 
   beam line in the transverse plane (impact parameter), $d \leq$ 0.3 cm; 
   (4) the $z$-vertex match $\leq$ 5 cm;
   (5) the distance between the extrapolated track and the track segment
   in the muon chamber, $\Delta x = r \Delta \phi \leq$ 2 cm.

   We select events containing at least one electron with $E_T\geq 8\; \gev$
   or one muon with $p_T \geq 8\; \gevc$. This selection produces a data
   sample quite similar to that used for the measurement of the 
   $B^{0}-\bar{B}^0$ flavor oscillation~\cite{mixing}. Since we are 
   interested in semileptonic decays of heavy quarks, trigger leptons are 
   also required to be non-isolated; we require $I \geq 0.1$, where the 
   isolation $I$ is defined as the ratio of the additional transverse 
   energy deposited in the calorimeter in a cone of radius 
   $R = \sqrt{\delta \phi^2 + \delta \eta^2} = 0.4$ around the lepton 
   direction to the lepton transverse energy.

   Further selection of the data sample is based upon jet reconstruction.
   Jets are reconstructed from the energy deposited in the calorimeter using
   a clustering algorithm with a fixed cone of radius $R=0.4$. A detailed
   description of the algorithm can be found in Ref.~\cite{jet_clus}. 
   Jet energies can be mismeasured for a variety of reasons (calorimeter 
   non-linearity, loss of low momentum particles because of the magnetic 
   field, contributions from the underlying event, out-of-cone losses, 
   undetected energy carried by muons and neutrinos). Corrections, which
   depend on the jet $E_T$ and $\eta$, are applied to jet energies; they 
   compensate for these mismeasurements on average but do not improve the 
   jet energy resolution. In this analysis we select central jets (taggable)
   by requiring that they include at least two SVX tracks~\cite{svx}.

   We require the trigger lepton to be contained in a cone of radius $R=0.4$
   around the axis of a taggable jet with uncorrected transverse energy
   $E_T \geq 15\; \gev$. This jet will be referred to as lepton-jet or 
   $e$-jet or $\mu$-jet. We also require the presence of at least one 
   additional taggable jet (away-jet) with $E_T \geq 15\; \gev$. The 
   requirement of a non-isolated lepton inside a jet rejects most of the
   leptonic decays of vector bosons and the Drell-Yan contribution. The 
   request of two jets with $E_T \geq 15\; \gev$ reduces the statistics of
   the data sample~\footnote{
   A jet with uncorrected transverse energy $E_T=15\; \gev$ corresponds
   to a parton with average transverse energy $<E_T>\simeq 20\; \gev$.}.
   This $E_T$-threshold is chosen because efficiencies and backgrounds of
   the SECVTX, JPB and SLT algorithms have been evaluated only for jets 
   with transverse energy above this value~\cite{cdf-tsig}. We select
   $68544$ events with an $e$-jet and $14966$ events with a $\mu$-jet.

   In order to determine the bottom and charmed content of the data we use
   two algorithms (SECVTX and JPB) which have been studied in detail in
   Refs.~\cite{suj,cdf-tsig}. SECVTX is based on the determination of the
   primary event vertex and the reconstruction of additional secondary
   vertices using displaced SVX tracks contained inside jets. Jet-probability
   (JPB) compares track impact parameters to measured resolution functions
   in order to calculate for each jet a probability that there are no
   long-lived particles in the jet cone~\cite{jpb}.
 
   The simulation of these tagging algorithms makes use of parametrizations
   of the detector response for single tracks, which were derived from the
   data. Because of the naivety of the method, these algorithms have required 
   several empirical adjustments. SECVTX tags not produced by hadrons with 
   heavy flavor (mistags) are underestimated by the detector simulation.
   Therefore SECVTX and JPB mistags are evaluated using a parametrized 
   probability derived from generic-jet data~\cite{cdf-tsig}, and are
   subtracted from the data in order to compare to the heavy flavor 
   simulation. We estimate that the mistag removal has a 10\%
   uncertainty~\cite{cdf-tsig}. 

   The tagging efficiency of these algorithms is not well modeled by the
   parametrized simulation. In Ref.~\cite{cdf-tsig}, we have used generic
   jets and a subset of the inclusive electron sample to determine the
   data-to-simulation scale factors for the tagging efficiency of these
   algorithms. The data-to-simulation scale factor of the SECVTX tagging
   efficiency for $b$-jets is measured to be $1.25\pm 0.08$. The number of
   tags in the simulation is multiplied by this scale factor, and we add 
   a 6\% uncertainty to the prediction of tags. The data-to-simulation scale
   factor for $c$ jets, has been measured to be $0.92\pm 0.28$~\cite{cdf-tsig};
   because of its large uncertainty, this scale factor is not implemented
   into the simulation, but we add a 28\% uncertainty to the prediction of
   tags due to $c$ jets. The data-to-simulation scale factor for the 
   jet-probability algorithm has been measured to be $0.96\pm 0.05$. 
   The number of tags in the simulation is multiplied by this scale factor, 
   and we add a $6$\% uncertainty to the prediction of tags.  

   In this study, we also probe the heavy-quark contribution by searching a
   jet for soft leptons ($e$ and $\mu$) produced by the decay of hadrons
   with heavy flavor. The soft lepton tagging algorithm is applied to sets
   of CTC tracks associated with jets with $E_T \geq 15\; \gev$ and 
   $|\eta| \leq $2.0. CTC tracks are associated with a jet if they are inside
   a cone of radius 0.4 centered around the jet axis. In order to maintain
   high efficiency, the lepton $p_T$ threshold is set low at $2\; \gevc$.
   To search for soft electrons the algorithm extrapolates each track to the
   calorimeter and attempts to match it to a CES cluster. The matched CES
   cluster is required to be consistent in shape and position with the 
   expectation for electron showers. In addition, it is required that  
   $0.7\leq E/p \leq 1.5$ and $E_{had}/E_{em}\leq 0.1$. The track specific
   ionization ($dE/dx$), measured in the CTC, is required to be consistent
   with the electron hypothesis. The efficiency of the selection criteria 
   has been determined using a sample of electrons produced by photon
   conversions~\cite{cdf-evid}.

   To identify soft muons, track segments reconstructed in the CMU, CMP
   and CMX systems are matched to CTC tracks. The CMU and CMX systems are 
   used to identify muons with $2\leq p_T\leq 3\; \gevc$ and 
   $p_T\geq 2\; \gevc$, respectively. Muon candidate tracks with 
   $p_T\geq 3\; \gevc$ within the CMU and CMP fiducial volume are
   required to match to track segments in both systems. The reconstruction
   efficiency has been measured using samples of muons from
   $J/\psi \rightarrow \mu^{+}\mu^{-}$ and $Z \rightarrow \mu^{+}\mu^{-}$
   decays~\cite{cdf-evid}.

   In the simulation, SLT tags are defined as tracks matching at generator
   level electrons or muons originating from $b$- or $c$-hadron decays 
   (including those coming from $\tau$ or $\psi$ cascade decays). The SLT
   tagging efficiency is implemented in the simulation by weighting these 
   tracks with the efficiency of each SLT selection criteria measured using 
   the data. The uncertainty of the SLT efficiency is estimated to be 
   10\% and includes the uncertainty of the semileptonic branching 
   ratios~\cite{cdf-evid,kestenb}.

   Rates of fake SLT tags are evaluated using a parametrized probability, 
   $P_{f}$, derived in special samples of generic-jet data, and are 
   subtracted from the data. This parametrization has been derived from the
   probability $P$ that a track satisfying the fiducial requirements produces
   an SLT tag. This probability is computed separately for each lepton flavor
   and detector type and is parametrized as a function of the transverse 
   momentum and isolation of the track~\cite{cdf-evid,kestenb}.
   In Ref.~\cite{cdf-tsig}, by fitting the impact parameter distributions
   of the SLT tracks in the same generic-jet samples used to derive the $P$ 
   parametrization, we have estimated that $P_f =(0.740 \pm 0.074) \times P$.
   It follows that, in generic-jet data, the probability that a track
   corresponds to a lepton arising from heavy-flavor decays is 
   $P_{hf}=(0.260 \pm 0.074) \times P$. Since we search a jet for SLT 
   candidates in a cone of radius of 0.4 around its axis, the probabilities
   of finding a fake SLT tag in a jet is 
   $P_{f}^{jet}(N) = \sum_{i=1}^{N} (1-P_{f}^{jet}(i-1)) \times P_f^i $, 
   where $N$ is the number of tracks contained in the jet cone.
   In generic jets, the probability of finding an SLT tag due to heavy flavor
   is  $P_{hf}^{jet}(N)=\sum_{i=1}^{N}(1-P_{hf}^{jet}(i-1))\times P_{hf}^i$.
   In Ref.\cite{cdf-tsig}, the uncertainty of the 
   $P^{jet}=P_{hf}^{jet}+P_{f}^{jet}$ parametrization has been estimated
   to be no larger than 10\% by comparing its prediction to the number of 
   SLT tags observed in 7 additional generic-jet samples.

   The efficiency for finding supertags (SLT tags in jets with SECVTX or 
   JPB tags) in the simulation is additionally corrected with a 
   data-to-simulation scale factor, $0.85\pm 0.05$, derived in a previous
   study of generic-jet data~\cite{suj}. The number of simulated supertags
   is multiplied by this factor, and we add a 6\% uncertainty to the 
   prediction of supertags. As mentioned earlier, the simulation of the 
   SLT algorithm uses parametrized efficiencies measured using samples of
   electrons from photon conversions and muons from 
   $J/\psi \rightarrow \mu^{+}\mu^{-}$ and $Z \rightarrow \mu^{+}\mu^{-}$ 
   decays. Since these leptons are generally more isolated than leptons from
   heavy flavor decays, we have some evidence that the efficiency of the SLT
   algorithm in the simulation is overestimated. However, since a reduced
   efficiency for finding supertags could also be generated by a reduced
   efficiency of the SECVTX (JPB) algorithm in jets containing a soft lepton,
   we have chosen to correct the simulated efficiency for finding supertags,
   but not the efficiency of the simulated SLT algorithm~\cite{suj}.
  \section{Data sample composition}
  \label{sec:ss-comp}
   The lepton-jets in our sample come from three sources: $b\bar{b}$ 
   production, $c\bar{c}$ production, and light quark or gluon production 
   in which a hadron mimics the experimental signature of a lepton 
   (fake lepton). The yield of fake leptons in light jets returned by our
   detector simulation cannot be trusted, and the $b\bar{b}$ and $c\bar{c}$
   production cross sections have large experimental and theoretical
   uncertainties. Therefore, we use measured rates of lepton-jets with SECVTX
   and JPB tags due to heavy flavor (i.e. after mistag removal) in order
   to separate the fractions of lepton-jets due to $b\bar{b}$ production 
   and $c\bar{c}$ production. The simultaneous use of the two tagging
   algorithms was pioneered in Ref.~\cite{cdf-tsig}; it allows to separate
   the $b$- and $c$-quark contributions because both algorithms have the 
   same tagging efficiency for $b$ jets, while for $c$ jets the efficiency
   of the JPB algorithm is approximately 2.5 times larger than that of the 
   SECVTX algorithm. The $b$ and $c$ content of away-jets is also determined
   with this method.

   The heavy flavor content of away-jets recoiling against a lepton-jet with
   heavy flavor depends on the production mechanisms (LO terms yield higher 
   fractions of heavy flavor than NLO terms). Therefore, we tune the
   cross sections of the various production mechanisms predicted by the 
   simulation to reproduce the observed number of lepton- and away-jets 
   with SECVTX and JPB tags due to heavy flavor.
 
   The fraction $F_{hf}$ of lepton-jets due to heavy flavor, before tagging,
   is estimated using the tuned simulation. The remaining fraction, 
   ($1-F_{hf}$), of lepton-jets is attributed to fake leptons in light jets. 
   The number of tags in away-jets, which recoil against a lepton-jet without
   heavy flavor, is predicted as $N_{a-jet}\times (1-F_{hf})\times P_{GQCD}$,
   where $N_{a-jet}$ is the total number of away-jets, and $P_{GQCD}$ is the
   average probability of tagging away-jets that recoil against lepton-jets
   without heavy flavor. The average probability $P_{GQCD}$ is estimated by
   weighting all the away-jets with a parametrized probability of finding
   SECVTX (or JPB) tags due to heavy flavor in generic-jet data
   ~\cite{cdf-tsig}. The number $N_{a-jet} \times (1-F_{hf}) \times P_{GQCD}$
   is subtracted from the number of tagged away-jets with heavy flavor that
   are used to tune the simulation. In Ref.~\cite{cdf-tsig}, this method has
   been cross-checked by using it also in a sample of data in which electrons
   are identified as coming from photon conversions. The heavy-flavor purity
   of $e$-jets due to photon conversions ($\simeq 8$\%) is depleted with
   respect to that of e-jets not due to conversions ($\simeq 50$\%). 
   The study in Ref.~\cite{cdf-tsig} shows that the usage of the probability
   $P_{GQCD}$ allows us to model the observed rate of tagged away-jets in 
   both the electron and conversion samples within a 10\% statistical 
   uncertainty. Therefore we attribute a 10\% uncertainty to the average
   probability $P_{GQCD}$.
  \subsection {Simulation of heavy flavor production and decay}
  \label{sec:ss-simulat}
   We use the {\sc herwig} Monte Carlo generator~\footnote{
   We use option 1500 of version 5.6, generic $2 \rightarrow 2$ hard 
   scattering with $p_T\geq 13\; \gevc$ (see Appendix~A in Ref.~\cite{suj}
   for more details).}
   to describe the fraction of data in which the lepton-jets contain hadrons
   with heavy flavor. We use the MRS(G) set of parton distribution functions
   \cite{mrsg}, and set $m_c = 1.5\; \gevcc$ and $m_b = 4.75\; \gevcc$.
   In the generic hard parton scattering, $b\bar{b}$ and $c\bar{c}$ pairs
   are generated by {\sc herwig} through processes of order $\alpha_s^{2}$
   such as $gg \rightarrow b\bar{b}$ (direct production). Processes of order
   $\alpha_s^{3}$  are implemented in the generator through flavor excitation
   processes, such as $gb \rightarrow g b$, or gluon splitting, in which the
   process $gg \rightarrow gg$ is followed by $g \rightarrow b\bar{b}$. The
   {\sc herwig} generator neglects virtual emission graphs, but, as all parton
   shower Monte Carlo generators, also includes higher than NLO diagrams.

   The bottom and charmed hadrons produced in the final state are decayed
   using the CLEO Monte Carlo generator ({\sc qq})~\cite{cleo}. At this 
   generation level, we retain only final states which contain hadrons with
   heavy flavor and at least one lepton with $p_T \geq 8\; \gevc$. The 
   accepted events are passed through a simulation of the CDF detector
   ({\sc qfl}) that is based on parametrizations of the detector response
   derived from the data. After the simulation of the CDF detector, the 
   Monte Carlo events are treated as real data. The simulated inclusive
   electron sample has $27136$ events, corresponding to a luminosity
   of $98.9\; {\rm pb}^{-1}$. The simulated inclusive muon sample has 
   $7266$ events, corresponding to a luminosity of $55.1\; {\rm pb}^{-1}$.
   The simulated samples have approximately the same luminosity as the data.
  \section{Determination of the rates of SECVTX and JPB tags due to
           heavy flavor in the data}\label{sec:ss-datatag}
   The heavy flavor content of the data is estimated from the number of jets
   tagged with the SECVTX and JPB algorithms. The numbers of lepton-jets and
   away-jets in the data, $N_{l-jet}$ and $N_{a-jet}$, are listed in
   Table~\ref{tab:datatag_1}. $N_{l-jet}$ is equal to the number of events
   and $N_{a-jet}$ is about 10\% larger, which means that about 10\% of the
   events have two away-jets. This table lists the following numbers of tags
   due to the presence of hadrons with heavy flavor:
  \begin{enumerate}
  \item  $T_{l-jet}^{SEC}$ and $T_{l-jet}^{JPB}$, the number of lepton-jets
         with a SECVTX and JPB tag, respectively.
  \item  $T_{a-jet}^{SEC}$ and $T_{a-jet}^{JPB}$, the number of away-jets
         with a SECVTX and JPB tag, respectively.
  \item  $DT^{SEC}$ and $DT^{JPB}$, the number of events in which the 
  	 lepton-jet and one away-jet are both tagged by SECVTX and JPB, 
 	 respectively.
  \end{enumerate}
   The uncertainty on the number of tags due to heavy flavor in 
   Table~\ref{tab:datatag_1} includes the 10\% error of the mistag removal. 

   Events in which the lepton-jet does not contain heavy flavor are not
   described by the heavy flavor simulation. In these events, the number of
   away-jets with tags due to heavy flavor is predicted using the average 
   tagging probabilities $P_{GQCD}$ listed in Table~\ref{tab:datatag_1}.
   These probabilities are used to correct the numbers of tagged away-jets
   that will be used to tune the heavy flavor simulation.
  \newpage
  \input{datatag_1.tex}
  \section{Tagging rates in the simulation} \label{sec:ss-simtag}
   Numbers of tags in simulated events which contain heavy flavor (h.f.), 
   characterized by the prefix $HF$, are listed in Table~\ref{tab:simtag_1}.
 
   Different production mechanisms are separated by inspecting at generator 
   level the flavor of the initial and final state partons involved in the 
   hard scattering. We attribute to flavor excitation the events in which 
   at least one of the incoming partons has heavy flavor and to direct
   production the events in which the incoming partons have no heavy flavor
   and the outgoing partons both have heavy flavor. Pairs of heavy quarks 
   which appear at the end of the evolution process are attributed to gluon
   splitting. The flavor type of each simulated jet is determined by
   inspecting its hadron composition at generator level.
  \newpage
  \input{tab_simtag_1.tex}

  \section{Tuning of the SM~simulation using SECVTX and JPB tags}
  \label{sec:ss-norm}
   Following the procedure outlined in Sec.~\ref{sec:ss-comp}, we fit the 
   data with the heavy flavor simulation using rates of jets before and 
   after tagging with the SECVTX and JPB algorithms. In the fit, we tune the
   cross sections of the different flavor production mechanisms. Starting
   from Table~\ref{tab:simtag_1} the simulated rate of jets before tagging
   can be written as: 
  \begin{eqnarray*}
    HF_{l,i} & = & K_{l} \cdot (HF_{b-dir, l,i} +
                  bf \cdot HF_{b-f.exc,l,i} + bg \cdot HF_{b-gsp,l,i} ) + \\
             &   & K_{l} \cdot (c \cdot HF_{c-dir,l,i} + 
          	  cf \cdot HF_{c-f.exc,l,i} + cg \cdot HF_{c-gsp,l,i} ) 
  \end{eqnarray*}
   The rates of tagged jets  are:
  \begin{eqnarray*}
    HFT^{j}_{l,i} & = & K_{l} \cdot SF_b^{j} (HFT^{j}_{b-dir,l,i} +
              	       bf \cdot HFT^{j}_{b-f.exc,l,i} +
                       bg \cdot HFT^{j}_{b-gsp,l,i}) + \\
   		 &   & K_{l} \cdot SF_c^{j} (c \cdot HFT^{j}_{c-dir,l,i} + 
             	       cf \cdot HFT^{j}_{c-f.ex,l,i} + 
             	       cg \cdot HFT^{j}_{c-gsp,l,i} )
  \end{eqnarray*}
   and the rates of events with a double tag are:
  \begin{eqnarray*}
    HFDT^{j}_{l}  & = & K_{l} \cdot {SF_b^{j}}^{2} 
            	       (HFDT^{j}_{b-dir,l} + bf \cdot HFDT^{j}_{b-f.ex,l} + 
            	       bg\cdot HFDT^{j}_{b-gsp,l}) + \\
   	  	&   & K_{l}\cdot {SF_c^{j}}^{2} (c \cdot HFDT^{j}_{c-dir,l} + 
            	    cf\cdot HFDT^{j}_{c-f.ex,l} + cg \cdot HFDT^{j}_{c-gsp,l}) 
  \end{eqnarray*}
   where the index $l$ indicates electron or muon data, $i$ indicates the 
   lepton- or the away-jet, and $j$ indicates the type of tag (SECVTX or JPB).
   The fit parameters $K_{l}$ account for the slightly different luminosity
   between data and simulation; they also include the normalization of the
   direct $b$-production cross section. The factors $c,cf,cg,bf$ and $bg$ 
   are fit parameters used to adjust the remaining cross sections calculated 
   by {\sc herwig} with respect to the direct $b\bar{b}$ production.
   The number of tags predicted by the simulation is obtained by multiplying
   the numbers in Table~\ref{tab:simtag_1} by the appropriate scale factor.
   The fit parameters $SF_b^{j}$ and $SF_c^{j}$ are used to account for the
   uncertainties of the corresponding scale factors. The simulated rates 
   $HFT^{j}_{l,i}$ and $HFDT^{j}_{l}$ have statistical errors
   $\delta^{j}_{T,l,i}$ and $\delta^{j}_{DT,l}$.

   As mentioned at the end of Sec.~\ref{sec:ss-comp}, the fraction of the
   data, which contains heavy flavor and is described by the simulation, is
   $F_{hf}^l=HF_{l,l-jet}/N_{l-jet}$. Therefore we fit the simulated rates
   to the quantities
  \begin{eqnarray*}
    HFT^{j}_{l,l-jet}({\rm DATA}) & = & T^{j}_{l,l-jet}  \\
    HFT^{j}_{l,a-jet}({\rm DATA}) & = & T^{j}_{l,a-jet} - N_{l,a-jet} \cdot
 			                (1- F_{hf}^l) \cdot P^{j}_{GQCD,l}\\
    HFDT^{j}_{l}({\rm DATA})      & = & DT^{j}_{l}
  \end{eqnarray*}
   where $P^{j}_{GQCD,l}$ is the probability of finding a type-$j$ tag due 
   to heavy flavor in a-jets recoiling against a l-jet without heavy flavor 
   (see Table~\ref{tab:datatag_1}). The errors $\epsilon^{j}_{T,l,i}$ of the
   rates $HFT^{j}_{l,i}({\rm DATA})$ include also the 10\% uncertainty of 
   $P^{j}_{GQCD,l,i}$.

   Following the same procedure pioneered in Ref.~\cite{cdf-tsig}, in which
   the {\sc herwig} simulation was tuned to generic-jet data, we constrain 
   the following fit parameters $X_i$ to their measured or expected value 
   $\bar{X}_i$ using the term 
   \[ G_i =  \frac { ( X_i - \bar{X}_i )^{2} } { \sigma^{2}_{\bar{X}_i} } \]
  \begin{enumerate}
  \item the ratio of the $b$ and $c$ direct production cross sections; it is 
        constrained to the {\sc herwig} default value with a 14\% Gaussian 
        error to account for the uncertainty of the parton fragmentation and
        for the fact that all quarks are treated as massless by the generator.
  \item the ratio of the $b$ to $c$ flavor excitation cross sections; it is 
        constrained to the {\sc herwig} default value with a 28\% error to
        account for the uncertainty of the parton structure functions.
  \item the correction $bg$ to the rate of gluon splitting; 
        $g \rightarrow b\bar{b}$ is constrained to the value $1.4 \pm 0.19$  
        returned by the fit to generic-jet data~\cite{cdf-tsig}.
  \item the correction $cg$ to $g \rightarrow c\bar{c}$; it is constrained to 
        the value $1.35 \pm 0.36$ returned by the fit to generic-jet 
        data~\cite{cdf-tsig}.
  \item we constrain $SF_b$ for SECVTX to unity with a 6\% error.
  \item we constrain $SF_c$ to unity with a  28\% error.
  \item we constrain $SF^{JPB}_b$ and $SF^{JPB}_c$ to unity with a 6\% error.
  \end{enumerate}
    In summary the fit minimizes the function
  \footnotesize
  \begin{eqnarray*}
   \chi^2 & = & \sum_{l= e,\mu} \sum_{j={\rm tag-type}} \left( 
		\sum_{i={\rm jet-type} } 
                \frac {( HFT^{j}_{l,i}({\rm DATA})- HFT^{j}_{l,i})^2 }
                {{\delta^{j}_{T,l,i}}^{2} +{\epsilon^{j}_{T,l,i}}^{2} }
                + \frac {( HFDT^{j}_{l}({\rm DATA})- HFDT^{j}_{l})^2 }  
                {{\delta^{j}_{DT,l}}^{2} +{\epsilon^{j}_{DT,l}}^{2} }
		\right) + \sum_{i=1}^{7} G_i
  \end{eqnarray*}
  \normalsize
   In total we fit 12 rates with 10 free parameters and 7 constraints.
   The best fit returns a $\chi^2$ value of 4.6 for 9 degrees of freedom.  
   The values of the parameters returned by the fit and their correlation
   coefficients are shown in Tables~\ref{tab:norm_1} and~\ref{tab:norm_1a}.
   Tagging rates in the data and in the fitted simulation are listed in
   Table~\ref{tab:tab_norm_2}.

   As shown by Table~\ref{tab:norm_1}, the correction factors to the 
   parton-level cross sections predicted by {\sc herwig} are close to unity.
   As also noted in Ref.~\cite{field}, {\sc herwig} predicts an inclusive 
   $b$-quark cross section at the Tevatron which is approximately a factor
   of two larger than the NLO prediction~\cite{mnr} and is in fair
   agreement with the CDF and D${\not \! {\rm O}}$~measurements. 
   As shown in Table~\ref{tab:simtag_1}, LO (labeled as direct production)
   and higher order (labeled as flavor excitation and gluon splitting) terms
   produce events with quite different kinematics. The LO contribution mostly
   consists of events which contain two jets with $b$ (or $c$) flavor in the
   detector acceptance. In contrast, only a small fraction of the events due
   to higher order terms contains two jets with heavy flavor in the detector
   acceptance. Therefore, the observed ratio of tagged a-jets to tagged 
   l-jets constrains the relative weight of LO and higher order contributions.
   In the {\sc herwig} simulation tuned to reproduce the data, the 
   contribution of higher order terms is approximately a factor of three
   larger than the LO contribution. The NLO prediction, which uses  
   normalization and factorization scales $\mu_0 = ({{p_T}_b}^2+m_b^2)^{1/2}$,
   underestimates the heavy flavor cross section by a factor of two and also
   yields LO and NLO contributions of approximately the same size; the tuned
   parton-level prediction of {\sc herwig} indicates that the data would be
   better described by a NLO calculation that uses the renormalization
   scale $\mu_r \simeq 0.5 \times ({{p_T}_b}^2+m_b^2)^{1/2}$ and the
   factorization scale $\mu_f \simeq 0.1 \times ({{p_T}_b}^2+m_b^2)^{1/2}$.

   As shown by the comparison between data and tuned simulation
   in Table~\ref{tab:tab_norm_2} (rows 3 to 6), the number of events 
   containing two jets with heavy flavor, corresponding to $\sigma_{b\bar{b}}$,
   is well modeled by the {\sc herwig} generator in which, as shown in
   Table~\ref{tab:simtag_1}, approximately 30\% of the production is due to
   higher-than-LO terms. Therefore the NLO prediction of $\sigma_{b\bar{b}}$ 
   underestimates the data by 20\%, whereas, as mentioned in the introduction,
   the NLO predictions of $\sigma_{b\bar{b}} \times BR$ and 
   $\sigma_{b\bar{b}} \times BR^2$ underestimate the data by a much larger 
   factor.
 \input{tab_norm_1.tex}

 \clearpage
 \input{tab_norm_1a.tex}

 \input{tab_norm_2.tex}
 \clearpage
  \subsection{Kinematics}~\label{sec:ss-smkin}
\begin{figure}[htp]
 \begin{center}
 \vspace{-0.4in}
 \leavevmode
 \epsfxsize \textwidth
 \epsffile{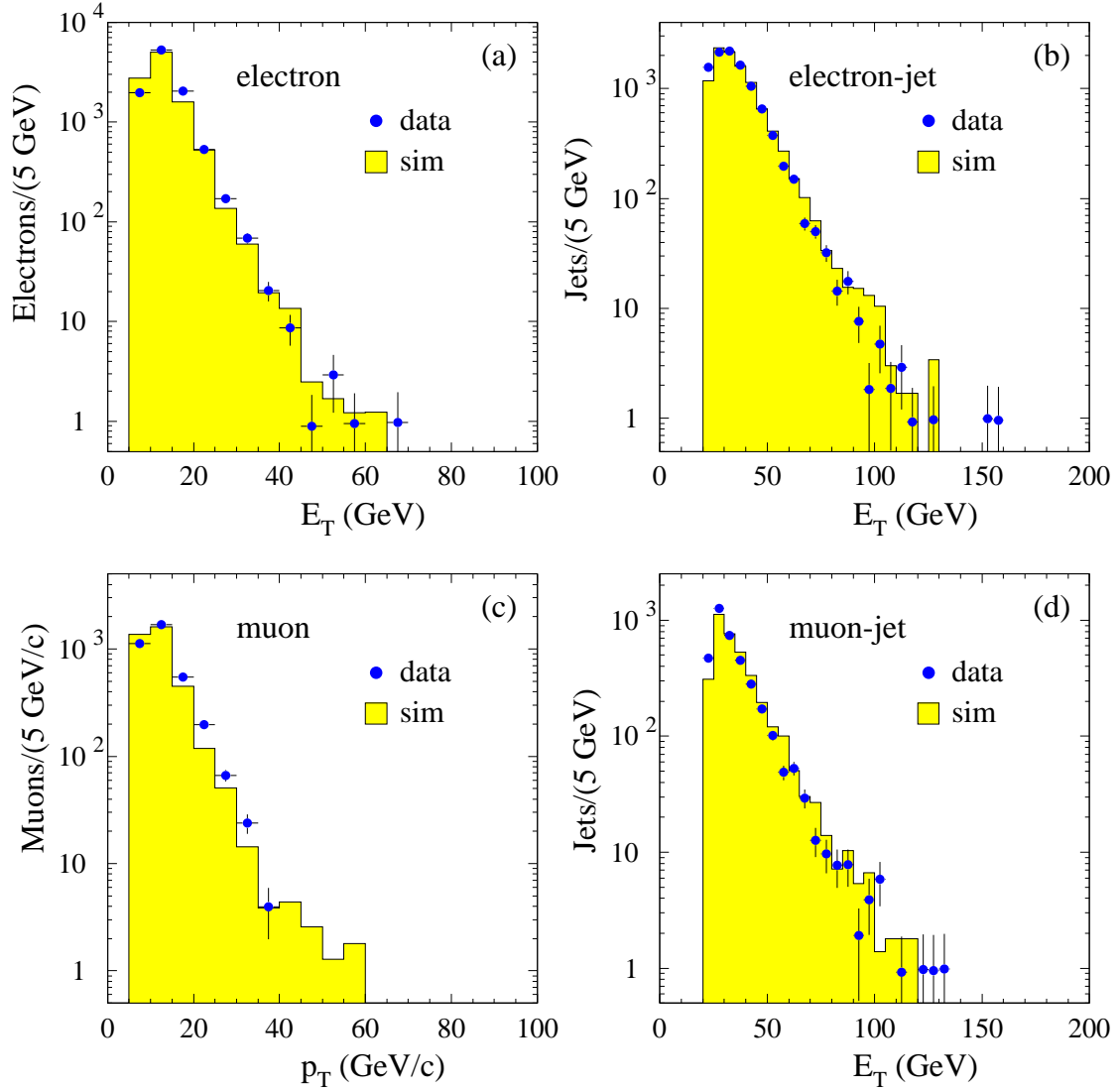}
 \caption[]{Distributions of transverse energy, $E_T$, or momentum, $p_T$, 
            for lepton-jets tagged by SECVTX.
            (a): electrons; 
            (b): electron-jets; 
            (c): muons; 
            (d): muon-jets. Jet energies are corrected for detector effects
            and out-of-cone losses.}
 \label{fig:fig1_smkin}
 \end{center}
 \end{figure}

\begin{figure}
 \begin{center}
 \leavevmode
 \epsfxsize \textwidth
 \epsffile{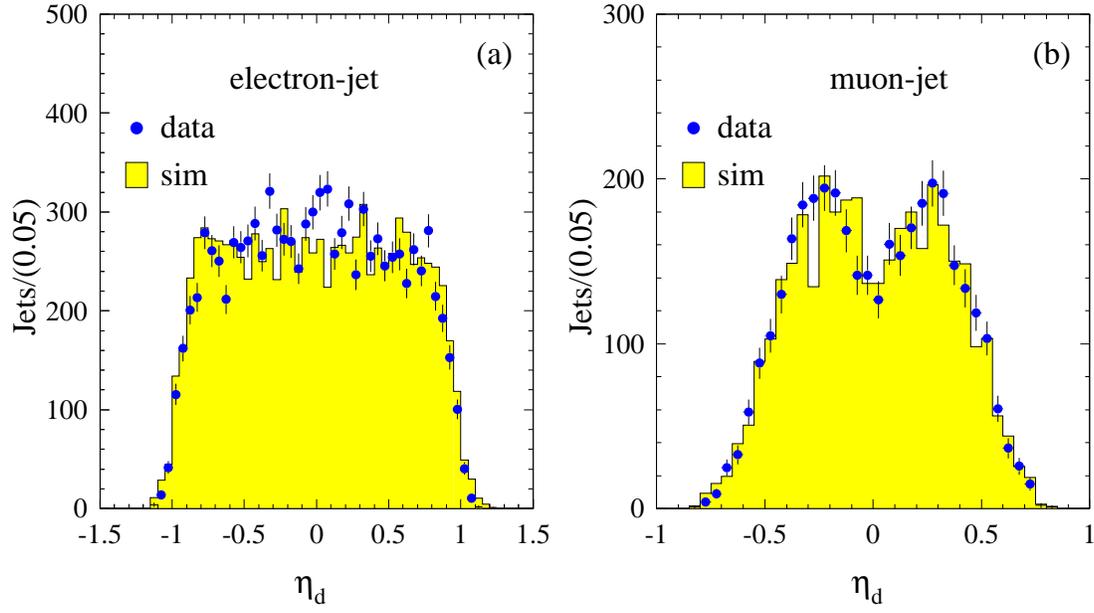}
 \caption[]{Pseudo-rapidity distributions of electron (a) and muon (b) jets 
            tagged by SECVTX. }
 \label{fig:fig2_smkin}
 \end{center}
 \end{figure}

\begin{figure}
 \begin{center}
 \leavevmode
 \epsfxsize \textwidth
 \epsffile{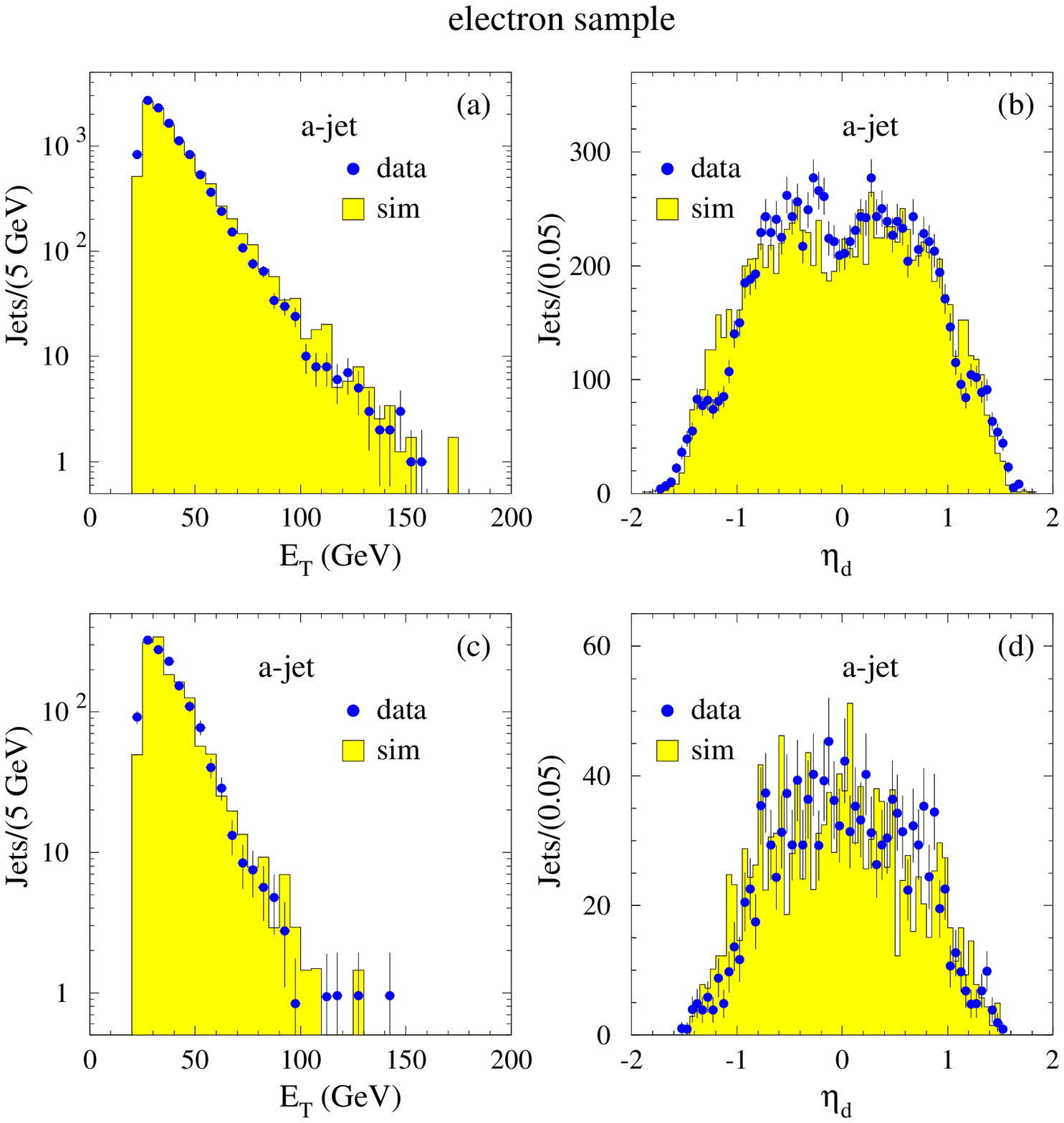}
 \caption[]{Away-jet distributions in events where the electron-jet is tagged
            by SECVTX.
            (a): a-jet transverse energy;
            (b): a-jet pseudo-rapidity;
            (c): transverse energy of a-jets tagged by SECVTX;
            (d): pseudo-rapidity of a-jets tagged by SECVTX. Jet energies are
            corrected for detector effects and out-of-cone losses.}
 \label{fig:fig3_smkin}
 \end{center}
 \end{figure}

\begin{figure}
 \begin{center}
 \leavevmode
 \epsfxsize \textwidth
 \epsffile{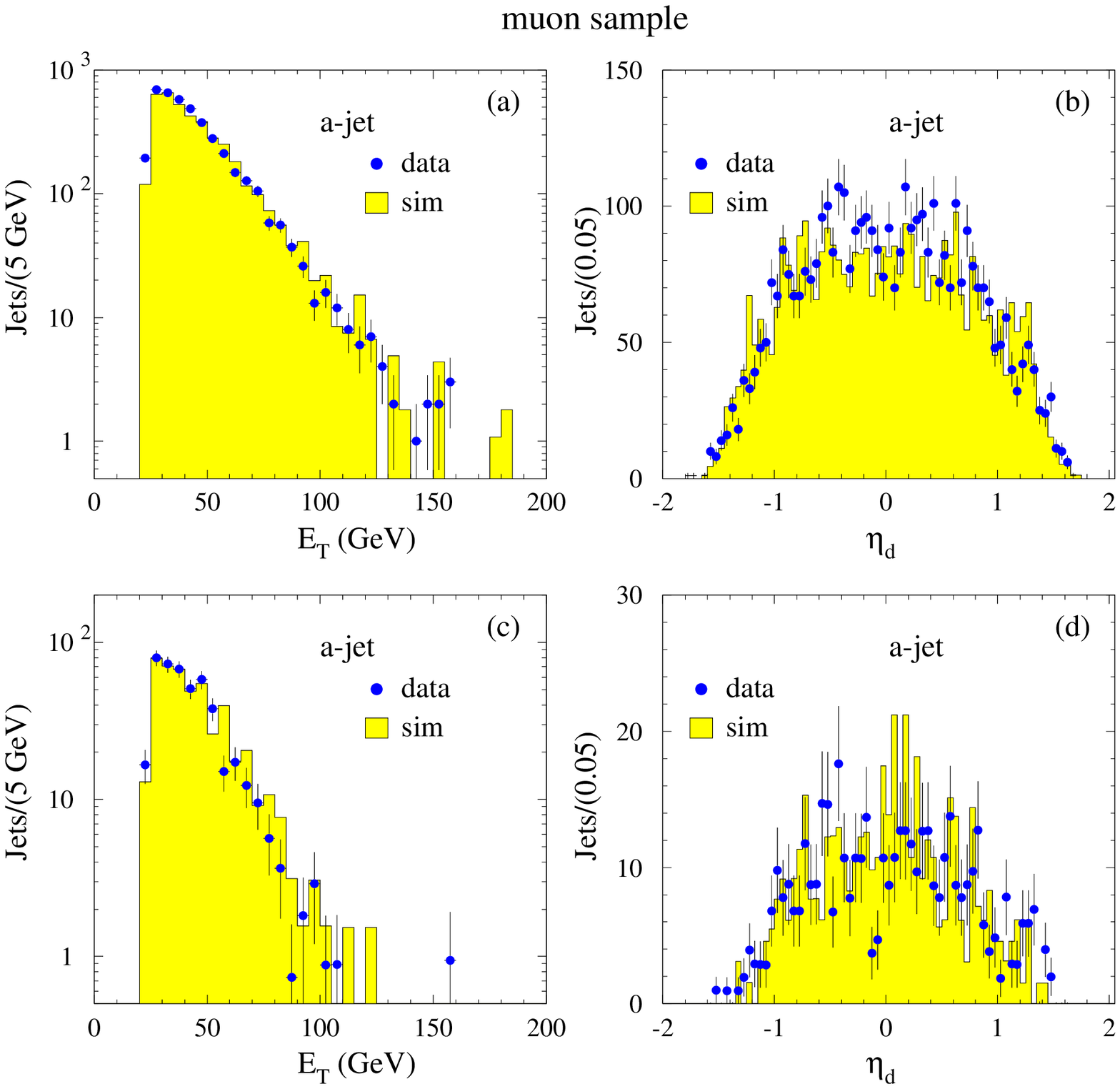}
 \caption[]{Away-jet distributions in events where the muon-jet is tagged
            by SECVTX.
            (a): a-jet transverse energy;
            (b): a-jet pseudo-rapidity;
            (c): transverse energy of a-jets tagged by SECVTX;
            (d): pseudo-rapidity of a-jets tagged by SECVTX. Jet energies are
            corrected for detector effects and out-of-cone losses.}
 \label{fig:fig4_smkin}
 \end{center}
 \end{figure}

   Because of the large flavor excitation contribution, the cross section 
   evaluated with {\sc herwig} depends strongly on the pseudo-rapidity and
   transverse momentum of the heavy quarks in the final state.
   The $2 \rightarrow 2$ hard scattering with $p^{\rm min}_T \geq 13\; \gevc$
   used to generate simulated events does not cover some of the available
   phase space, such as the production of massive gluons with small 
   transverse momentum, which then branch into pairs of heavy quarks.
   In addition, the detector simulation ({\sc qfl}), which is based upon
   parametrizations of single particle kinematics, may not accurately model
   the jet-$E_T$ and trigger thresholds used in the analysis. It is therefore
   important to show that the simulation, which  reproduces correctly the
   tagging rates and the away-jet multiplicity distribution, also models the
   event kinematics. Figures~\ref{fig:fig1_smkin}~to~\ref{fig:fig4_smkin}
   compare transverse energy and pseudo-rapidity distributions in the data
   and in the the simulation, normalized according to the fit listed in
   Table~\ref{tab:norm_1}~\footnote{
   The systematic discrepancy in the first bin of each $E_T$ distribution is
   the reflection of the slightly inaccurate modeling of the efficiency of
   the lepton trigger near the threshold. A few local discrepancies in some
   pseudo-rapidity distributions at  $|\eta| \simeq 0$ and $|\eta|\simeq 1$ 
   are due to an inaccurate modeling of the calorimetry cracks. These small
   discrepancies are not relevant in this analysis.}.

\begin{figure}
 \begin{center}
 \leavevmode
 \epsffile{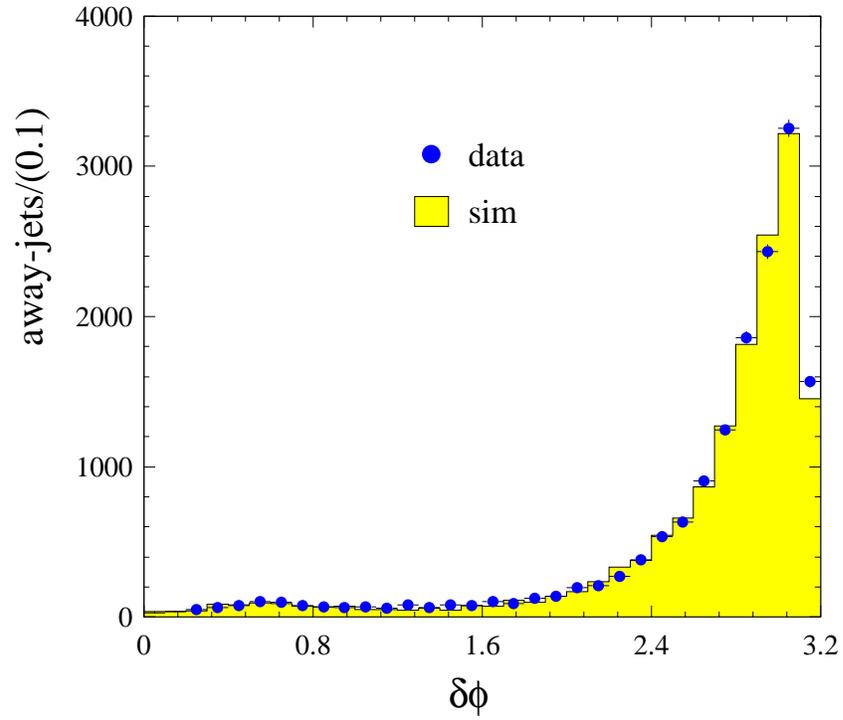}
 \caption[]{Distribution of the azimuthal angle $\delta\phi$ between
            lepton-jets tagged by SECVTX and away-jets in the same event.}
 \label{fig:fig5_smkin}
 \end{center}
 \end{figure}

   Figure~\ref{fig:fig5_smkin} compares distributions of the azimuthal angle
   $\delta \phi$ between the lepton-jet and the away-jets. The region at 
   $\delta \phi$ smaller than 1.2, which is well modeled by the tuned
   simulation, is mostly populated by the gluon splitting contribution. The
   good agreement between data and prediction supports the 40\% increase of
   the gluon splitting cross sections (see Table~\ref{tab:norm_1}).  

\begin{figure}
 \begin{center}
 \leavevmode
 \epsfxsize \textwidth
 \epsffile{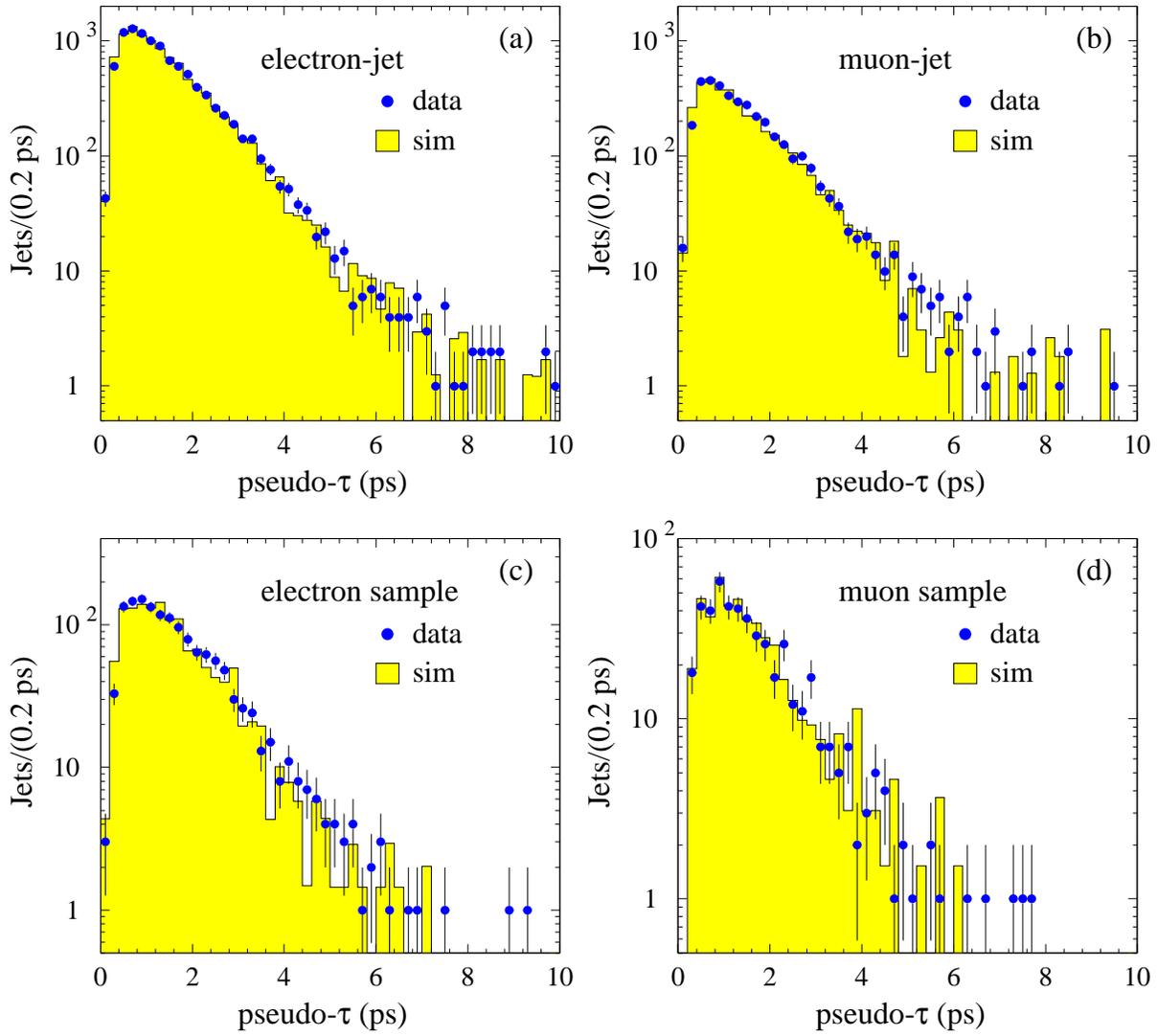}
 \caption[]{Pseudo-$\tau$ distributions of electron-jets (a) and muon-jets
            (b) tagged by SECVTX and for tagged away-jets in events where
            the electron-jet (c) or the muon-jet (d) is also tagged.}
 \label{fig:fig6_smkin}
 \end{center}
 \end{figure}

   Figure~\ref{fig:fig6_smkin} compares pseudo-lifetime distributions of 
   SECVTX tags. The pseudo-lifetime is defined as
   \[ {\rm pseudo-}\tau = \frac{L_{xy} \cdot M^{SVX}}{c \cdot p_{T}^{SVX}} \]
   where $L_{xy}$ is the projection of the two-dimensional vector pointing
   from the primary vertex to the secondary vertex on the jet direction, 
   and $M^{SVX}$ and $p_{T}^{SVX}$ are the invariant mass and the transverse 
   momentum of all tracks forming the SECVTX tag. 
 
\begin{figure}
 \begin{center}
 \leavevmode
 \epsfxsize \textwidth
 \epsffile{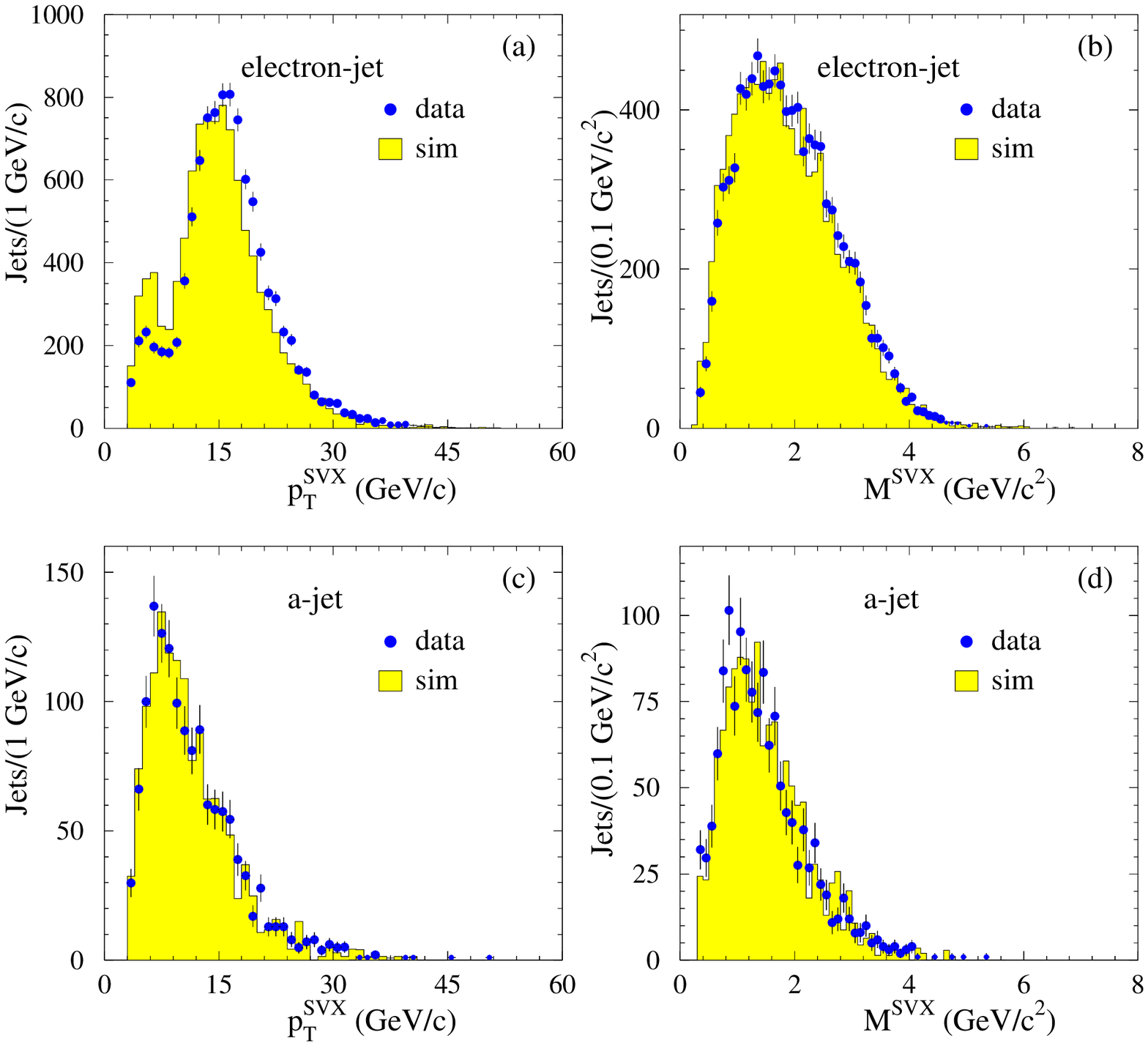}
 \caption[]{Distributions of the transverse momentum (a) and invariant mass
            (b) of SECVTX tags in electron-jets; (c) and (d) are analogous  
            distributions for away-jets in events in which the e-jet is also 
            tagged.}
 \label{fig:fig7_smkin}
 \end{center}
 \end{figure}

\begin{figure}
 \begin{center}
 \leavevmode
 \epsfxsize \textwidth
 \epsffile{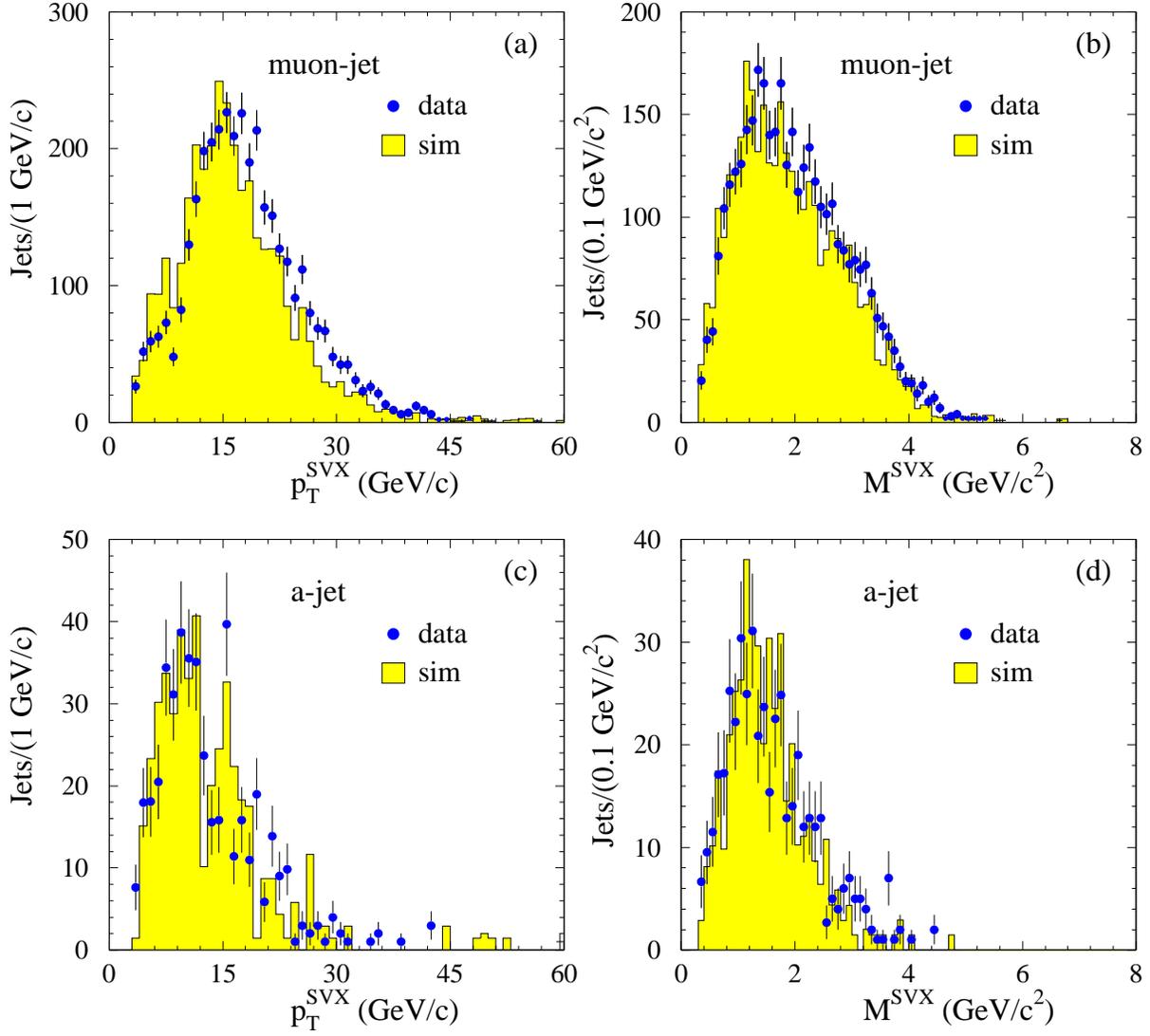}
 \caption[]{Distributions of the transverse momentum (a) and invariant mass
            (b) of SECVTX tags in muon-jets; (c) and (d) are analogous 
            distributions for away-jets in events in which the muon-jet is 
	    also tagged.}
 \label{fig:fig8_smkin}
 \end{center}
 \end{figure}

   Distributions of $M^{SVX}$ and $p_{T}^{SVX}$, which is sensitive to the
   heavy quark fragmentation, are shown in Figures~\ref{fig:fig7_smkin} 
   and~\ref{fig:fig8_smkin}. In Figures~\ref{fig:fig7_smkin}(a) 
   and~\ref{fig:fig8_smkin}(a), the simulated $p_{T}^{SVX}$ distributions of
   SECVTX tags in lepton-jets are above the data near to the $p_T$-threshold.
   This discrepancy follows from the fact that the tagging efficiency in the
   simulation is smaller than in the data and we take care of it with an
   overall multiplicative factor. This procedure does not account for the
   fact that the probability that a $8\; \gevc$ lepton is part of a tag is
   also higher in the data than in the simulation. In away-jets, where
   high-$p_T$ tracks are not a selection prerequisite, there is better
   agreement between data and simulation. 

   In conclusion, our simulation calibrated within the theoretical and
   experimental uncertainties models correctly the heavy flavor production 
   at the Tevatron. 
 \newpage
 \newpage
 \newpage
 \newpage
  \section{Rates of SLT tags}~\label{sec:ss-rateslt}
   Following the strategy outlined in Sec.~\ref{sec:strat}, we search 
   away-jets for soft leptons ($e$ or $\mu$) with $p_T \geq 2\; \gevc$ and
   contained in a cone of radius 0.4 around the jet axis. We then compare
   rates of away-jets containing soft lepton tags due to heavy flavor in 
   the data and in the simulation  tuned as in Table~\ref{tab:norm_1}.
   Table~\ref{tab:tab_rateslt_1} lists the following rates of away-jets
   with SLT tags:
  \begin{enumerate}
  \item $T_{a-jet}^{SLT}$, the number of away-jets with a soft lepton tag.
  \item $T_{a-jet}^{SLT \cdot SEC}$ ($T_{a-jet}^{SLT \cdot JPB}$), the
        number of away-jets with an SLT tag and a SECVTX (JPB) tag (called
        supertag in Ref.~\cite{suj}).
  \end{enumerate}
   The uncertainty on the number of tags due to heavy flavor in 
   Table~\ref{tab:tab_rateslt_1} includes the 10\% error of the mistag 
   removal. In events in which the lepton-jet does not contain heavy flavor, 
   the number of away-jets with an SLT tag due to heavy flavor is predicted
   using the average probability $P_{GQCD}$. This average probability is 
   estimated by weighting all the away-jets with the parametrized probability
   $P_{hf}^{jet}$, derived in generic-jet data and described
   in Sec.~\ref{sec:ss-sample}. In these events, the uncertainty of the 
   average probability of finding a real or a fake SLT tags is estimated to
   be no larger than 10\%. We cross-check the estimate of these uncertainties
   in Sec.~\ref{sec:ss-syst}. 

   Rates of SLT tags in the simulation before tuning are shown in
   Table~\ref{tab:tab_rateslt_2}. The uncertainty of the SLT efficiency is
   estimated to be 10\% and includes the uncertainty of the semileptonic  
   branching ratios~\cite{cdf-evid,kestenb}. The numbers of supertags
   predicted by the simulation are obtained by multiplying the numbers in
   Table~\ref{tab:tab_rateslt_2} by the scale factor $0.85\pm 0.05$.

   Following the notations of Section~\ref{sec:ss-norm}, rates of tagged 
   away-jets with heavy flavor in the fitted simulation are defined as:
  \begin{eqnarray*}
    HFT^{SLT}_{l,a-jet} & = & K_{l} \cdot (HFT^{SLT}_{b-dir,l,a-jet} +
                              bf \cdot HFT^{SLT}_{b-f.exc,l,a-jet} +
                              bg \cdot HFT^{SLT}_{b-gsp,l,a-jet}) + \\
                        & = & K_{l} \cdot (c \cdot HFT^{SLT}_{c-dir,l,a-jet} +
                              cf \cdot HFT^{SLT}_{c-f.exc,l,a-jet} +
                              cg \cdot HFT^{SLT}_{c-gsp,l,a-jet} ) 
	 	 	      \mbox{~~~~~and~} \\
   HFT^{SLT\cdot j}_{l,a-jet} & = & 
          K_{l} \cdot SF_b^{j} (HFT^{SLT \cdot j}_{b-dir,l,a-jet} +
          bf \cdot HFT^{SLT \cdot j}_{b-f.exc,l,a-jet} +
          bg \cdot HFT^{SLT \cdot j}_{b-gsp,l,a-jet}) + \\
   & &    K_{l} \cdot  SF_c^{j} (c \cdot HFT^{SLT \cdot j}_{c-dir,l,a-jet} +
          cf \cdot HFT^{SLT \cdot j}_{c-f.exc,l,a-jet} +
          cg \cdot HFT^{SLT \cdot j}_{c-gsp,l,a-jet} )
  \end{eqnarray*}
   where $HFT^{SLT}_{l,a-jet}$ is the rate of a-jets containing heavy flavor
   tagged by the SLT algorithm, and $HFT^{SLT\cdot j}_{l,a-jet}$ is the rate
   of a-jets containing heavy flavor with a supertag $j$ (SECVTX or JPB).
   The errors on the simulated rates include the statistical error,
   the systematic uncertainty for finding SLT tags and supertags, and the
   uncertainties of the parameters ($ K_{l}$, $bf$, $bg$, $c$, $cf$, $cg$,
   and $SF$) listed in  Table~\ref{tab:norm_1} and~\ref{tab:tab_norm_2}.
   In the data the analogous rates are:
 \begin{eqnarray*}
   HFT^{SLT}_{l,a-jet}({\rm DATA})& = & T^{SLT}_{l,a-jet} - N_{l,a-jet}
       \cdot (1-F_{hf}^l)\cdot P^{SLT}_{GQCD,l}  \mbox{~~~~~and~} \\
   HFT^{SLT \cdot j}_{l,a-jet}({\rm DATA})& = & T^{SLT \cdot j}_{l,a-jet} 
       - N_{l,a-jet} \cdot (1-F_{hf}^l)\cdot P^{SLT \cdot j}_{GQCD,l} 
 \end{eqnarray*}
 \newpage
 \input{tab_ratslt_1.tex}
 \input{tab_ratslt_2.tex}
  \subsection{Rates of soft leptons due to heavy flavor in the data
              and in the tuned simulation} \label{sec:ss-sltres}
   The comparison of the yields of away-jets with SLT tags due to heavy
   flavor in the data and in the tuned simulation is shown in
   Table~\ref{tab:tab_rateslt_5}. Table~\ref{tab:tab_rateslt_6} lists 
   the numbers of tags in the tuned simulation split by flavor type and
   production mechanism, and Table~\ref{tab:tab_rateslt_sum} summarizes
   the different contributions to the observed number of tags. In the data 
   there are $HF^{SLT}_{a-jet} = 1138\pm 140$ a-jets with a soft lepton 
   tag due to heavy flavor. The $\pm140$ error is dominated by the 10\%
   systematic uncertainty of the fake and generic-QCD contributions to SLT
   tags; the statistical error is $\pm51$ jets. The simulation predicts
   $747\pm75$ a-jets with soft lepton tags due to $b\bar{b}$ and 
   $c\bar{c}$ production (most of the error is systematic and due to the 
   10\% uncertainty on the SLT tagging efficiency). The discrepancy is a
   $2.5\; \sigma$ systematic effect.

   The comparison of the yields of supertags in the data and in the tuned
   simulation is also listed in Table~\ref{tab:tab_rateslt_sum}. 
   The subset of data, in which a-jets have both SLT and JPB tags due to 
   heavy flavor, contains $453 \pm 29$ supertags (in this case the 
   $\pm 25$ statistical error is larger than the $\pm 15$ systematic error
   due to the fake-tag subtraction). The simulation predicts $317\pm 25$ 
   a-jets with a supertag due to $b\bar{b}$ and $c\bar{c}$ production. 
   The $\pm 25$ systematic error is obtained combining in quadrature the
   uncertainty of the SLT efficiency ($\pm 16$) with the uncertainty
   ($\pm 20$) due to the fit in Table~\ref{tab:norm_1} and to the
   simulation statistical error. This discrepancy is a $3.5\; \sigma$ 
   effect dominated by systematic uncertainties. In the even smaller 
   subset of events, in which a-jets contain both SECVTX and SLT tags due
   to heavy flavor, the discrepancy between data and simulation is a 
   $2.4\; \sigma$ effect, also dominated by the same systematic errors.

   There is no gain in combining the three results because the uncertainties
   on the number of a-jets with SLT tags due to heavy flavor, before and 
   after tagging with the SECVTX and JPB algorithms, are highly correlated.
   Away-jets with supertags are a subset of the a-jets with SLT tags, and
   there is overlap between the subsets with JPB and SECVTX supertags. 
   However, it is important to note that the discrepancy between observed 
   and expected number of SLT tags is of the same size before and after 
   tagging with the SECVTX and JPB algorithms. This disfavors the 
   possibility that the disagreement between data and simulation arises 
   from jets containing hadrons with a lifetime much shorter than that of 
   conventional heavy flavor. 

\begin{figure}
 \begin{center}
 \leavevmode
 \epsffile{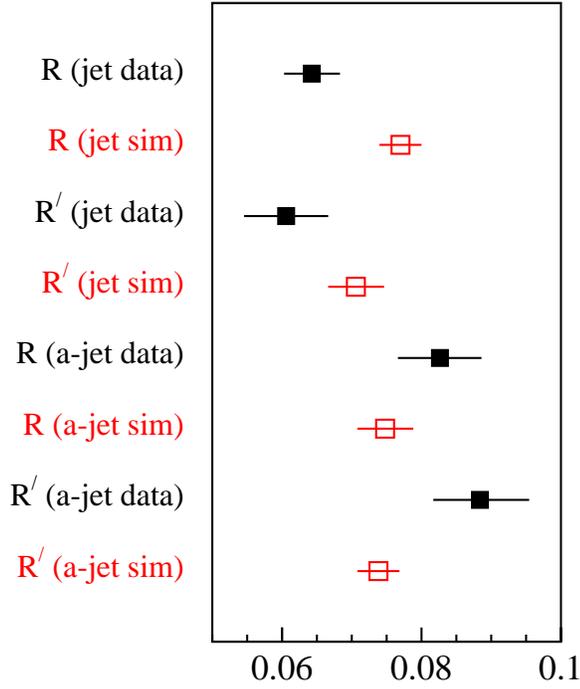}
 \caption[]{Yield of $R$, the ratio of the number of jets with a SECVTX and 
            SLT tag to that with a SECVTX tag in the data (square) and the
            corresponding simulations (open square). $R^{'}$ is the analogous
            ratio for JPB tags. The error in the simulation comes from the
            uncertainty of relative ratio of bottom and charmed hadron in the
            data; this uncertainty results from the tuning of the heavy flavor
            cross sections predicted by {\sc herwig} to model the rates of 
            SECVTX and JPB tags observed in the data. The simulation is not 
            corrected for the scale factor $0.85 \pm 0.05$ which is used to 
            equalize data and prediction in generic jets.} 
 \label{fig:fig_abw}
 \end{center}
 \end{figure}

   We have considered the impact on the number of expected supertags due to
   the $0.85 \pm 0.05$ scale factor derived in generic-jet data. If we had
   evaluated the number of simulated supertags using the product of 
   simulated efficiencies of the SECVTX (JPB) algorithm and of the SLT
   algorithm, which has a 10\% uncertainty, the discrepancy between data
   and simulation would be smaller: $1.6\; \sigma$ and $1.0\; \sigma$ for 
   a-jets with JPB and SECVTX tags, respectively. However, analogous rates
   of tags in generic-jet data would be approximately $1.5\; \sigma$ lower
   than in the simulation. Figure~\ref{fig:fig_abw} shows the yield of $R$,
   the ratio of the number of supertags (SECVTX+SLT) to that of SECVTX tags
   produced by heavy flavor, in generic jets and in the away-jets recoiling
   against a lepton-jet. The ratio $R^{'}$ is derived in analogy replacing
   SECVTX with JPB tags. The comparison of these ratios in the generic-jet
   data and their simulation has been used in Ref.~\cite{suj} to calibrate
   the efficiency for finding supertags in the simulation. In 
   Figure~\ref{fig:fig_abw}, the efficiency for finding supertags in the
   simulation has not been corrected with the $0.85 \pm 0.05$ scale factor.
   For the simulation, the plotted errors of $R\; (R^{'})$ account for the
   uncertainty of the relative contribution of $b$ and $c$ quarks, but not
   for the uncertainty of the supertag efficiency, which is no smaller than
   10\%. One notes that the simulation predicts the same value of $R\;
   (R^{'})$ for generic jets and away-jets in lepton-triggered events, 
   whereas, in the data, the value of $R\; (R^{'})$ for away-jets is
   approximately 20\% higher than for generic jets. 
 
   Finally, we have investigated the dependence of the predicted yield of 
   away-jets with SLT tags on the ratio of the $c\bar{c}$ to $b\bar{b}$ 
   productions predicted by the simulation. To a good approximation, the
   predicted yield does not depend on the tuning of the simulation. Since
   the ratio of the tagging efficiency for $c$ jets to that for $b$ jets
   is approximately equal for the JPB and SLT algorithms~\footnote{The 
   average tagging efficiencies in this data set are $\epsilon_b^{JPB}=0.43$,
   $\epsilon_c^{JPB}=0.30$, $\epsilon_b^{SLT}=0.064$, and 
   $\epsilon_c^{SLT}=0.046$.}, 
   the expected number of away-jets with SLT tags is
 \begin{eqnarray*}
   HFT_{a-jet}^{SLT} & = &\epsilon_b^{SLT} \times (N_b + \epsilon_c^{SLT}/ 
                          \epsilon_b^{SLT} \times N_c) =
     \epsilon_b^{SLT}/\epsilon_b^{JPB} \times \epsilon_b^{JPB} \times (N_b + 
     \epsilon_c^{JPB} /\epsilon_b^{JPB} \times N_c)  \\
   & = &  \epsilon_b^{SLT}/\epsilon_b^{JPB} \times HFT_{a-jet}^{JPB}(DATA)= 
    \epsilon_b^{SLT}/\epsilon_b^{JPB} \times (5126.6 \pm 146.7)=763 \pm 80
 \end{eqnarray*}
   and does not depend on the size of $N_b$ and $N_c$, the numbers of 
   away-jets attributed by the fit to bottom and charmed flavor, respectively.
   As an example of this, without constraining the ratio of the $c$ to $b$ 
   direct productions to the nominal value within a 14\% error, we have 
   misled the fit to return a very different, and not correct, local minimum 
   ($c=2.8 \pm 1.6$ instead of $c=1.01 \pm 0.10$ in Table~\ref{tab:norm_1}).
   The number of a-jets with SLT tags remains approximately constant 
   (in the electron sample, $598 \pm 69$ becomes $603 \pm 66$; 
   in the muon sample, $149 \pm 21$ becomes $156 \pm 21$). 
 \input{tab_ratslt_5n.tex}
 \input{tab_ratslt_6n.tex}
 \input{tab_ratslt_sum.tex}
 \clearpage
  \section{Systematics}\label{sec:ss-syst}
   This section reviews and verifies systematic effects that could reduce
   the discrepancy between observed and predicted numbers of away-jets with
   a soft lepton tag due to heavy flavor. The discrepancy depends on the
   estimate of the mistag rate in the data and on the simulated efficiency
   of the SLT algorithm, and also on the size of the $b\bar{b}$ contribution
   in the simulation. We cross-check these estimates in subsections~A and B,
   respectively. In subsection~C, we verify the discrepancy between data 
   and simulation found in this study with a sample of jets that recoil
   against $J/\psi$ mesons arising from $B$ decays.
  \subsection{Fake SLT tags and the simulated SLT efficiency} 
  \label{sec:fake-slt}
   Table~\ref{tab:tab_rateslt_sum} shows an excess of $391$ away-jets
   with SLT tags due to heavy flavor with respect to the number, 
   $747\pm75$, predicted by the heavy flavor simulation. In the data, 
   we have removed a fake contribution of $619 \pm 62$ SLT tags~\footnote{
   In the data, we have also subtracted the generic-jet contribution of
   SLT tags due to a-jets recoiling against l-jets without heavy flavor 
   (see Table~\ref{tab:tab_rateslt_sum}). This contribution is slightly
   overestimated because the tagging probability $P^{jet}$ has been
   constructed using also events in which both jets contain heavy flavor.}. 
   If the estimate of the fake rate could be increased by 60\% (6 times
   the estimated uncertainty), this excess would disappear. The simulated
   efficiency of the SLT algorithm has been tuned using the data and we
   estimate its uncertainty to be 10\%; however, if the simulated 
   efficiency could be increased by 50\%, the disagreement between data 
   and simulation would also disappear.

   Table~\ref{tab:tab_rateslt_sum} also shows an excess of $137$ a-jets 
   with SLT+JPB supertags due to heavy flavor with respect to the number 
   $316 \pm 25$ predicted by the simulation. In the data, we have removed
   $142\pm 14$ fake tags; in this case, one would need to increase the
   fake-rate estimate by  $10\; \sigma$ in order to cancel the excess
   in the data. The simulated supertag efficiency has been calibrated 
   with generic-jet data to a 6\% accuracy; in order to cancel the 
   discrepancy, the supertag efficiency in the simulation should be 
   increased by $8.7\; \sigma$.

   We verify the uncertainty of the fake rate and heavy flavor contributions
   by comparing rates of SLT tags in three generic-jet samples to their
   corresponding simulations fitted to the data using rates of SECVTX and
   JPB tags. These rates of tags, together with the fake contributions 
   evaluated with the same fake parametrizations used in the present study,
   are listed in Table~\ref{tab:tab_4.1}, which is derived from the study
   presented in Ref.~\cite{suj}. A summary of Table~\ref{tab:tab_4.1} is
   presented in Table~\ref{tab:tab_4.1bis}. The observed number of SLT tags
   in generic jets (sample A in Table~\ref{tab:tab_4.1bis}) is dominated
   by the fake contribution, and we use the difference between the observed
   number of SLT tags  and the number of SLT tags due to heavy flavor 
   predicted by the simulation to reduce the uncertainty of the fake rate.
   Generic-jet data contain $18885$ SLT tags. The parametrized probability
   predicts $15570 \pm 1557$ fake tags. The simulation predicts $3102$ SLT
   tags due to heavy flavor with a 13\% uncertainty (dominated by the 10\% 
   uncertainty of the SLT tagging efficiency). By removing from the data 
   the heavy flavor contribution predicted by the simulation, one derives
   an independent and consistent estimate for the fake contribution of
   $15783 \pm 403$ SLT tags. The latter determination of the fake
   contribution has a 2.6\% uncertainty.

   Before tagging with the SLT algorithm, away-jets in the inclusive lepton
   sample have a larger heavy flavor content ($\simeq 26$\%) than that of 
   sample~A in Table~\ref{tab:tab_4.1bis} ($\simeq 13$\%). However, generic 
   jets tagged by SECVTX and JPB algorithms (samples B and C, respectively)
   have a heavy-flavor purity of 78\% and 58\%, respectively. Because 
   these latter samples have a larger heavy flavor content, the discrepancy
   between the observed and predicted yields of away-jets with SLT tags
   observed in the present study cannot arise from deficiencies of the 
   heavy flavor simulation or from an increase of the fake probability 
   in jets with heavy favor. 
 
   In addition, the total number of SLT tags observed in generic jets can 
   be used to achieve a better determination of the sum of the predicted
   numbers of fake SLT tags plus SLT tags due to heavy flavor (h.f.) with 
   respect to that presented in Sec.~\ref{sec:ss-sltres}. To obtain this,
   we fit the observed rate of SLT tags in both samples~A and C with the 
   predicted number of fake and h.f. tags weighted with unknown parameters
   $P_f$ and $P_{h.f.}$, respectively. The data constrain the parameter 
   values to be $P_f = 1.017 \pm 0.013$ and $P_{h.f.} = 0.981 \pm 0.045$
   with a correlation coefficient $\rho = -0.77$.

   After having removed the contribution of events in which the lepton-jet
   does not contain heavy flavor, away-jets contain $1757 \pm 104$ SLT tags;
   in Sec.~\ref{sec:ss-sltres}, this number was compared to a prediction of 
   $619 \pm 62$ fake and $747 \pm 75$ h.f. tags. When using the weights, 
   errors and parameter correlation derived using generic jets, the 
   prediction of the total number of SLT tags becomes $1362 \pm 28$.
   The systematic uncertainty of the prediction is reduced by a factor 
   of $2.8$ with respect to that presented in Sec.~\ref{sec:ss-sltres}, 
   while the disagreement remains the same. In conclusion, the discrepancy 
   observed in this study cannot arise from obvious deficiencies of the
   prediction.
 \newpage
 \input{tab_4_1.tex}
 \newpage
 \input{tab_4_1bis.tex}
   We have investigated the possibility that the rate of fake SLT tags might
   be higher in jets with heavy flavor than in jets due to light partons.
   The correlation between the fake and h.f. predictions, established by
   the previous  comparison between the total number of observed and predicted
   tags in generic jets, would require that an increase of the fake rate is 
   compensated by a smaller efficiency of the SLT algorithm in the simulation, 
   and it would not reduce the disagreement between data and prediction
   observed in the inclusive lepton sample. However, it is of interest to 
   show this study in anticipation of the next subsection.

   The parametrization of the SLT fake rate has been derived in generic-jet
   data without distinguishing between muons faked by hadrons not contained
   by the calorimeter and muons produced by in-flight decays of $\pi$ and
   $K$ mesons. The second contribution is believed to be small because the
   reconstruction algorithms reject tracks which exhibit large kinks, but
   this has never been carefully checked. Away-jets in the inclusive lepton
   sample have a larger heavy flavor content ($\simeq 26$\%) than the generic
   jets used to determine the SLT fake rate ($\simeq 13$\%), and possibly
   a larger kaon content. Since kaons have a shorter lifetime than pions,
   in-flight decays of kaons could increase the SLT fake rate in the 
   inclusive lepton sample with respect to generic-jet data. We verify
   the contribution of kaon in-flight-decays by using a combination of data
   and simulation. First we extend the simulation of the SLT algorithm to
   match tracks not only to leptons originating from heavy quark decays at
   generation level but also to muons originating from kaon decays at 
   detector simulation level. With this implementation, the rate of SLT tags
   in the simulation increases by only 1\% (from $746.9$ to $754.4$ tags).

\begin{figure}[htp]
 \begin{center}
 \vspace{-0.2in}
 \leavevmode
 \epsffile{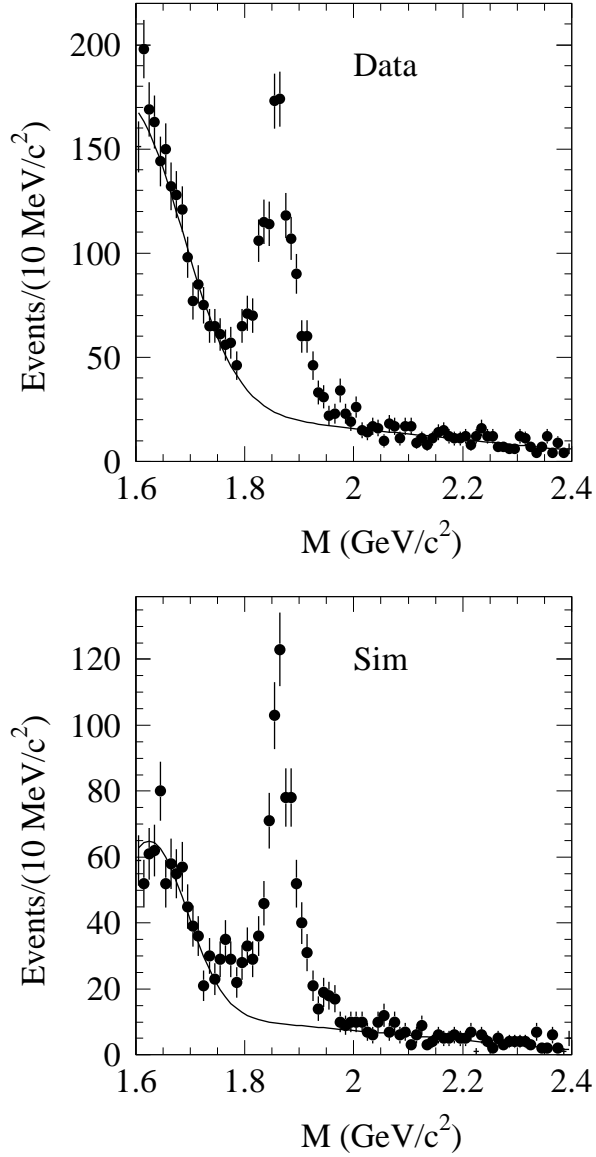}
 \caption[]{Distributions of the $K \pi$ invariant mass, $M$. The solid line
            is a polynomial fit to the distributions excluding the window
            between $1.7$ and $2.0\; \gevcc$.}
 \label{fig:fig_syst_1}
 \end{center}
 \end{figure}

   We check the simulation result within a factor of two by selecting 
   $D^{0} \rightarrow K \pi$ decays in the data and in the tuned simulation.
   As done in previous analyses~\cite{dkapi}, we search the inclusive lepton
   sample for $D^{0} \rightarrow K^{-} \pi^{+}$ decays near the trigger 
   leptons. To increase the sample statistics we do not require that leptons
   are contained in a jet with transverse energy larger than $15\; \gev$.
   The $D^{0} \rightarrow K^{-} \pi^{+}$ decays are reconstructed as follows.
   We select events in which a cone of radius 0.6 around the lepton direction
   contains only two SVX tracks with opposite charge, $p_T \geq 1.0\; \gevc$, 
   and an impact parameter significance larger than two~\footnote{
   The impact parameter is the distance of closest approach to the primary
   vertex in the transverse plane.}.
   We reconstruct the two-track invariant mass attributing the kaon mass to 
   the track with the same charge as the lepton as is the case in 
   semileptonic $B$-decays. The resulting $K^{-} \pi^{+}$ invariant mass 
   spectrum is shown in Figure~\ref{fig:fig_syst_1} together with a 
   polynomial fit to the background which ignores the mass region between
   $1.7$ and $2.0\; \gevcc$. According to the fit, in the mass range 
   $1.82 - 1.92\; \gevcc$ the simulation contains $563$ $D^{0}$ mesons
   on top of a background of $95$ events (the corresponding 563 kaons are
   also identified at generator level). We find that one kaon in $563$ 
   $D^{0}$ decays produces a soft muon tag, which corresponds to $0.0018$
   SLT tags per kaon. 

   The data contain $1117$ $K^{-} \pi^{+}$ pairs in the mass range 
   $1.82 - 1.92\; \gevcc$ ($891$ are attributed by the fit to $D^{0}$ 
   mesons and $226$ to the background). The $1117$ kaon tracks produce $6$ 
   SLT tags. The contribution of the background is estimated from the 
   side-bands ($1.64 - 1.74$ and $2.0 - 2.1\; \gevcc$) to be $3.8\pm 1.0$
   events. It follows that $891$ kaons from $D^{0}$ decays produce 
   $2.2\pm 2.6$ SLT tags. The fraction of SLT tags per kaon, 
   $0.0024\pm 0.0029$, includes the fake-tag contribution, and is 
   consistent with the small fraction predicted by the simulation. We 
   conclude that in-flight decays of $K$ mesons are a negligible background
   contribution. 
 \newpage
  \subsection{$b$ purity of the data sample} \label{sec:sec-bpur}
   The discrepancy between observed and predicted number of a-jets with SLT 
   tags due to heavy flavor would be reduced if the $b\bar{b}$ contribution
   was underestimated by the simulation. In this section, we verify that the
   $b\bar{b}$ contribution is predicted correctly. As shown in 
   Table~\ref{tab:tab_rateslt_6}, the inclusive electron simulation  
   predicts that 79\% of the away-jets with heavy flavor are due to 
   $b\bar{b}$ production. This table also shows that the fraction of  
   away-jets with an SLT tag is higher in events due to $b\bar{b}$ production
   (2\%) than in events due to $c\bar{c}$ production (1\%). If one had a
   reason to increase the $b$ purity in the simulation from 79\% to 100\%,
   one could increase the predicted number of a-jets with a SLT tag in 
   Table~\ref{tab:tab_rateslt_5} from $598$ to $756$, which is closer to the
   $865 \pm 115$ a-jets with a SLT tag due to heavy flavor in the data.
   We provide an independent check of the $b$ purity of the inclusive lepton
   sample by comparing the number of $D^{0}$, $D^{\pm}$, and $J/\psi$ mesons
   from $B$-decays which are contained in lepton-jets in the data and in the
   normalized simulation. 
  \subsubsection{$l^{-}D^{0}$ and $l^{+} D^{-}$ candidates}
   We identify $l^{-}D^{0}$ candidates searching for 
   $D^{0} \rightarrow K^{-} \pi^{+}$ decays inside the lepton-jet, as
   explained in the previous section. In a similar way, we identify
   $l^{+} D^{-}$ pairs searching for $D^{-} \rightarrow K^{+} \pi^{-}\pi^{-}$
   decays inside the lepton-jet. In this case, we select jets containing
   one positive and two negative tracks with $p_T \geq 0.6\; \gevc$ and
   impact parameter significance larger than 2.5 in a cone of radius 0.6
   around its axis. When reconstructing the three-track invariant mass, we
   attribute the kaon mass to the track with the same charge as the lepton
   as is the case in semileptonic $B$ decays.

\begin{figure}
 \begin{center}
 \leavevmode
 \epsfxsize \textwidth
 \epsffile{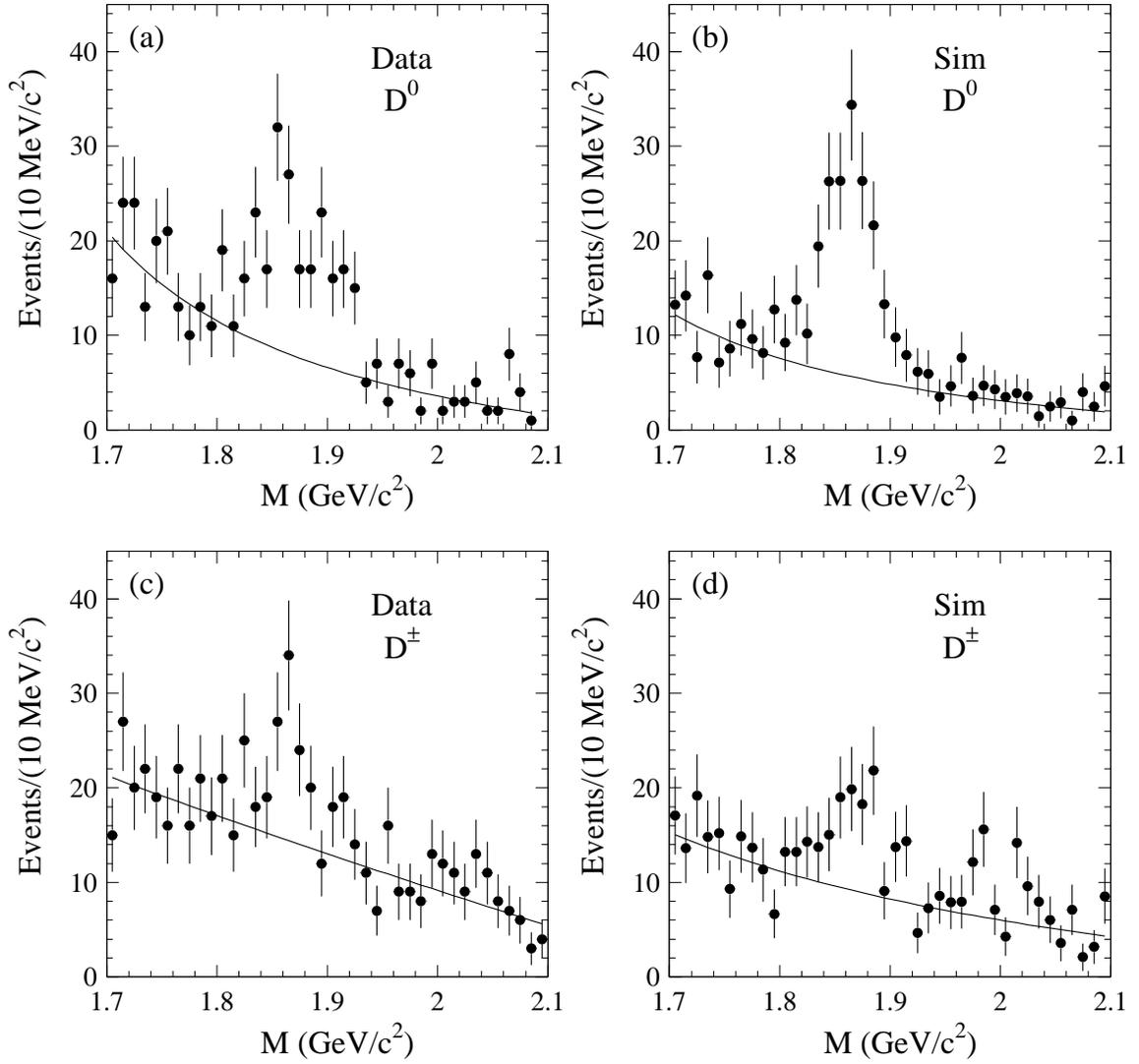}
 \caption[]{Invariant mass distributions of $D^{0}$ candidates in the data
            (a) and in the simulation (b) and of $D^{\pm}$ candidates in the
            data (c) and in the simulation (d). The solid line is a 
            polynomial fit to the mass distributions excluding the region 
            $1.75 - 2.0\; \gevcc$.}
 \label{fig:fig_syst_2}
 \end{center}
 \end{figure}

   Figure~\ref{fig:fig_syst_2} shows the invariant mass distributions of
   $D^{0}$ and $D^{\pm}$ candidates found in the data and in the fitted 
   simulation. By comparing with Figure~\ref{fig:fig_syst_1}, one notes that
   the mass resolution is degraded when using tracks inside a jet and is
   degraded slightly differently in the data and in the simulation.

   There are $83510$ lepton-jets in the data with an estimated heavy flavor
   purity $F_{hf} = (47.9 \pm 2.0)$\%. The simulation normalized according 
   to Table~\ref{tab:norm_1} contains $39989$ lepton-jets with heavy flavor.
   In the mass range $1.82 - 1.92\; \gevcc$, we find $205$ $D^{0}$ 
   candidates in the data and $195.5$ $D^{0}$ candidates in the simulation.
   By fitting the side-bands with a polynomial function (solid line in 
   Figure~\ref{fig:fig_syst_2}), we evaluate a background of $79.6 \pm 6.0$
   events in the data and of $55.6 \pm 5.5$ events in the simulation.
   After background subtraction, there are $126.0\pm 15.5$ $D^{0}$ mesons in 
   the data and $139.9 \pm 15.0$ $D^{0}$ mesons in the simulation.

   In the mass range $1.82 - 1.92\; \gevcc$, there are $216$ $D^{\pm}$
   candidates in the data and $159.2$ in the simulation. By fitting the 
   side-bands with a polynomial function we estimate a background of 
   $142.3 \pm 10.0$ events in the data and of $90.7 \pm 6.4$ events in the
   simulation. After background subtraction we find $73.7 \pm 17.8$ 
   $D^{\pm}$ mesons in the data and $68.5 \pm 14.1$ in the simulation. From
   the ratio of the numbers of $lD$ candidates, we derive that the ratio of
   the $b\bar{b}$ production in the simulation to that in the data is
   $1.09 \pm 0.15$.
  \subsubsection{$J/\psi$ candidates}
   We look for $J/\psi$ candidates by searching the electron- or muon-jet 
   for additional soft lepton tags with the same flavor and opposite charge. 
   Dileptons with invariant mass $2.6 \leq m_{ee} \leq 3.6\; \gevcc$ and
   $2.9 \leq m_{\mu \mu} \leq 3.3\; \gevcc$ are considered $J/\psi$ 
   candidates ($Dil_{\psi}$). $Dil^{SEC}$ and $Dil^{JPB}$ are the numbers
   of $J/\psi$ candidates in lepton-jets tagged by SECVTX and JPB, 
   respectively. We use the number of SS dileptons with a 10\% error to
   estimate and remove the background to OS dileptons due to misidentified
   leptons~\cite{drelly}.

\begin{figure}
 \begin{center}
 \leavevmode
 \epsfxsize \textwidth
 \epsffile{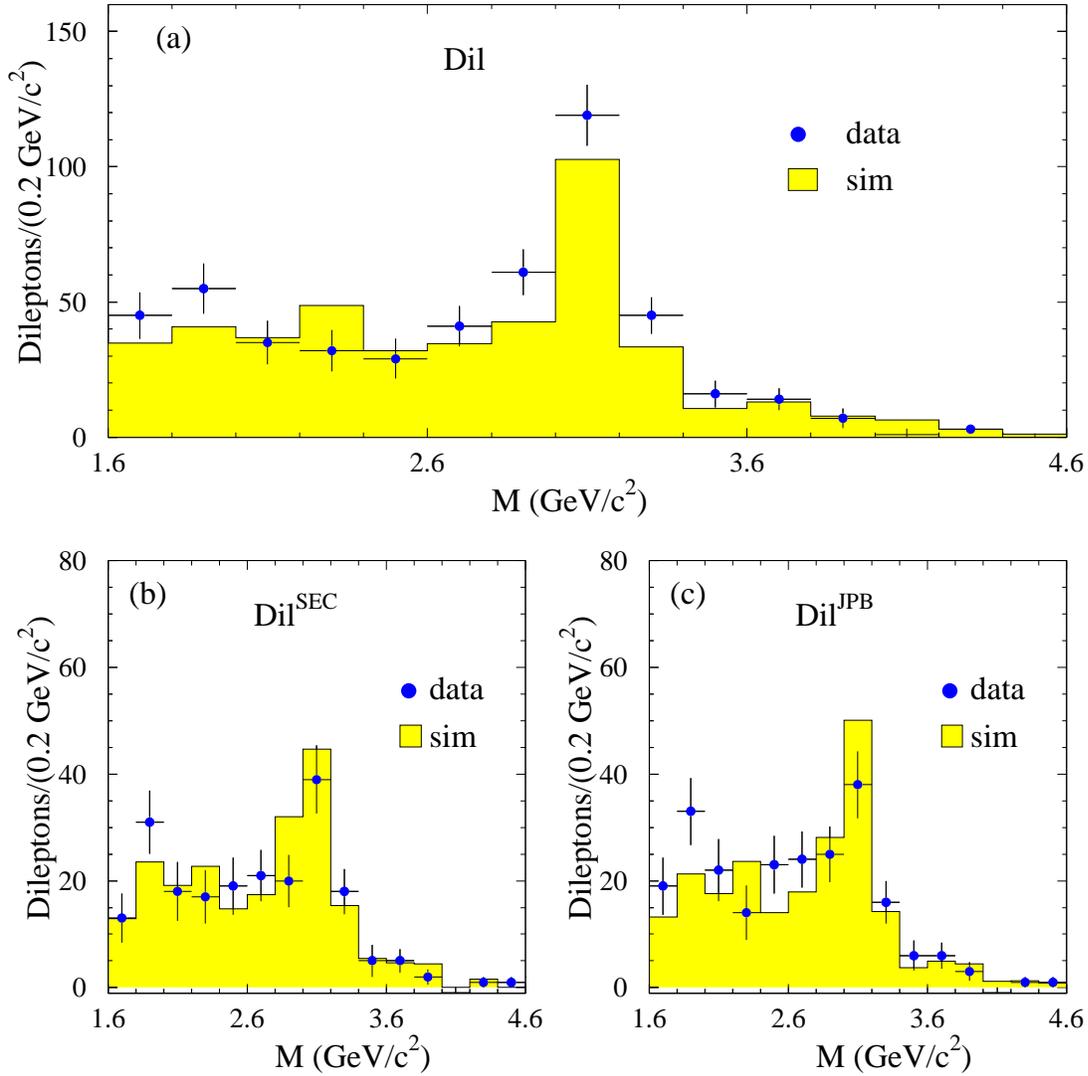}
 \caption[]{Distributions of the invariant mass of same flavor dileptons
            inside the same jet before (a) and after tagging with
            SECVTX (b) and JPB (c).}
 \label{fig:kinslt_5}
 \end{center}
 \end{figure}

   Figure~\ref{fig:kinslt_5} compares invariant mass distributions of same
   flavor dileptons including $J/\psi$ mesons in the data and in the 
   simulation (in the simulation $J/\psi$ mesons are only produced by $B$
   decays). Rates of $J/\psi$ mesons in the data and in the normalized
   simulation are listed in Table~\ref{tab:tab_rateslt_7}. One notes that 
   the simulation contains a number of $J/\psi$ mesons in jets tagged by
   SECVTX or JPB which is slightly higher than, but consistent with the data.
   Before tagging, the rate of $J/\psi$ mesons in the data is 20\% larger
   than in the simulation, whereas it was expected to be larger by a factor
   of two according to the CDF measurement of the fraction of $J/\psi$'s 
   coming from $B$-decays~\cite{psi-sec}. This would happen if the $b\bar{b}$
   cross section had been overestimated in normalizing the simulation.

   After combining the ratio of $lD$ candidates in the data to that in the
   simulation with the ratio of $l J/\psi$ candidates with a JPB tag listed 
   in Table~\ref{tab:tab_rateslt_7}, we estimate that the ratio of the 
   $b\bar{b}$ production in the simulation to that in the data is 
   $1.09 \pm 0.11$. This ratio is consistent with unity, and does not 
   support the possibility that the $b$ purity in the fitted simulation 
   is underestimated by 21\%.
 \input{tab_ratslt_7.tex}

 \clearpage
  \subsection{ $J/\psi \rightarrow \mu\mu$ data}  \label{sec:ss-psijet}
   As shown in Table~\ref{tab:tab_rateslt_6}, away-jets with a supertag
   are mostly due to $b\bar{b}$ production as it is the case for generic 
   jets with a supertag. However, we see a discrepancy between observed
   and predicted number of supertags after having calibrated the 
   supertag efficiency in the simulation by using generic jets. Since this
   is suggestive that the excess of SLT tags in the away-jets is related to
   the request that a jet contains a presumed semileptonic $b$-decay 
   (lepton-jet), we study a complementary data sample enriched in 
   $b\bar{b}$ production but not in semileptonic $b$-decays, i.e. events 
   containing $J/\psi \rightarrow \mu^{+}\mu^{-}$ decays. The data sample
   consists of $\simeq 110\; {\rm pb}^{-1}$ of $p\bar{p}$ collisions 
   collected by CDF during the $1992-1995$ collider run. This  sample has
   been used for many analyses and is described in detail in 
   Ref.~\cite{wenzel}. Approximately 18\% of these $J/\psi$ mesons come 
   from $B$ decays~\cite{psi-sec}. Muon candidates are selected as in
   Ref.~\cite{wenzel}. Since we want to make use of the $B$ lifetime to 
   remove the contribution of prompt $J/\psi$ mesons, we select muons with
   SVX tracks. The dimuon invariant mass is calculated without constraining
   the two muon tracks to a common vertex since the mass resolution is not
   important in this check. In addition we require a jet with transverse
   energy larger than $15\; \gev$ lying in the hemisphere opposite to the
   $J/\psi$ and contained in the SVX acceptance.

\begin{figure}
 \begin{center}
 \leavevmode
 \epsffile{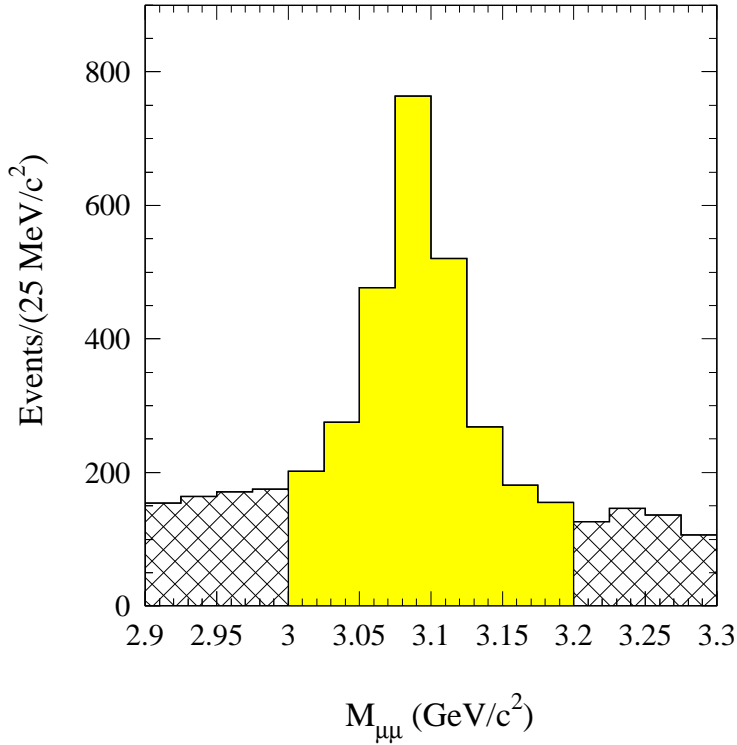}
 \caption[]{Invariant mass distribution of muon pairs. The shaded area 
            indicates the $J/\psi$ signal region and the cross-hatched area
            indicates the side-band region, SB, used to estimate the 
            background.}
 \label{fig:fig_syst_3}
 \end{center}
 \end{figure}

   The dimuon invariant mass distribution in these events is shown in 
   Figure~\ref{fig:fig_syst_3}. In the mass range between $3$ and 
   $3.2\; \gevcc$ there are $1163$ $J/\psi$ events over a background
   of $1179$ events estimated from the side-band region (see 
   Figure~\ref{fig:fig_syst_3})~\footnote{
   The request of a recoiling away-jet reduces the number of $J/\psi$
   mesons in the original data set by a factor of $\simeq 200$.}.

\begin{figure}
  \begin{center}
  \leavevmode
  \epsffile{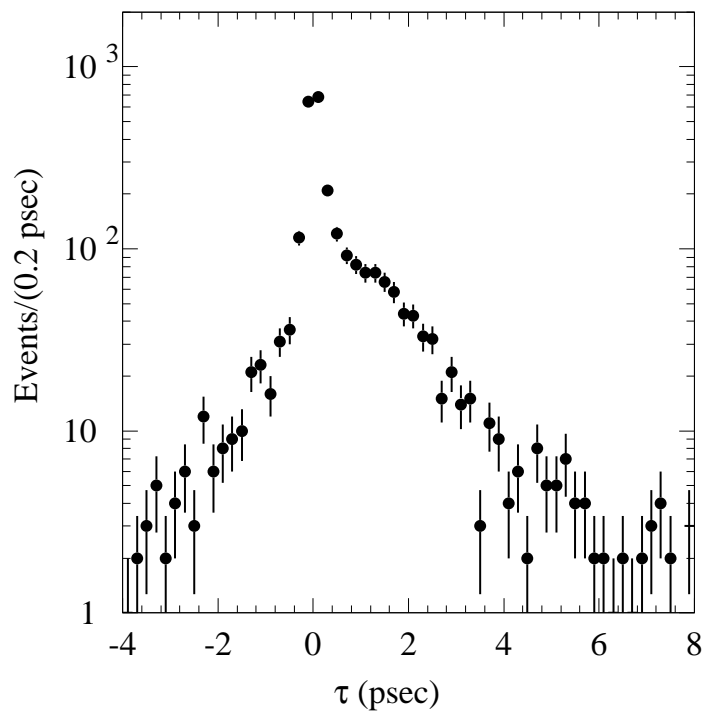}
  \caption[]{Lifetime distribution of $J/\psi$ candidates.}
  \label{fig:fig_syst_4}
  \end{center}
  \end{figure}

   The $J/\psi$ lifetime is defined as
   \[ \tau= \frac { (\vec{L}\cdot \vec{p}_T ) \cdot M} {c \cdot p_T^{2}} \]
   where $M$ and $p_T$ are the dimuon invariant mass and transverse momentum
   and $L$ is the distance between the event vertex and the origin of the
   muon tracks. The lifetime distribution of $J/\psi$ candidates is shown
   in Figure~\ref{fig:fig_syst_4}. As studied in Ref.~\cite{wenzel}, prompt
   $J/\psi$ candidates produce a symmetric $\tau$-distribution peaking
   at $\tau = 0$.  We call $\psi^{+}$ and $\psi^{-}$  the numbers of 
   $J/\psi$ candidates with positive and negative lifetime; $SB^{+}$ and
   $SB^{-}$ are the analogous numbers for the side-band region, which is
   used to estimate the background in the invariant mass distribution. The
   number of $J/\psi$ mesons from $B$ decays is then 
   $N_\psi= \psi^{+} - \psi^{-} - (SB^{+} - SB^{-}) = 561$ which is 48\% of 
   the initial sample. In the opposite hemisphere we find $572$ away-jets. 
   In these a-jets we measure the following numbers of tags after mistag
   removal:
  \begin{enumerate}
   \item $48.0 \pm 15.1$ SECVTX tags
   \item $61.7 \pm 17.3$ JPB tags
   \item $-9.4 \pm 14.4$ SLT tags
  \end{enumerate}
   For $54.8\pm 11.5$ lifetime tags (average of the observed number of 
   SECVTX and JPB tags) the simulation predicts $8.1\pm 1.7$ SLT tags. 
   The observed number of SLT tags is $1.2\; \sigma$ lower than the 
   prediction rather than $50$\% larger as in the inclusive lepton sample.
  \section{Conclusions}  \label{sec:concl}
   We have studied the heavy flavor properties of jets produced at the
   Tevatron collider. This study is motivated by the evidence, reported in
   Ref.~\cite{suj}, for a class of jets that contain long-lived objects
   consistent with $b$- or $c$-quark decays, identified by the presence of
   secondary vertices (SECVTX tags) or of tracks with large impact parameters
   (JPB tags), but which also have an anomalously large content of soft 
   leptons (SLT tags); we refer to these as superjets and supertags. The 
   study in Ref.~\cite{suj} focused on high-$p_T$ jets produced in 
   association with $\W$ bosons. The analysis reported here uses a much larger
   data set collected with low-$p_T$ lepton triggers ($p_T \geq\;8\; \gevc$).
   This data set has been previously used to study bottom and charmed 
   semileptonic decays, and to provide calibrations for the measurement of 
   the pair production of top quarks~\cite{cdf-tsig}.

   In the present analysis, we study events having two or more central jets
   with $E_T \geq 15\; \gev$, one of which (lepton-jet) is consistent with a
   semileptonic bottom or charmed decay to a lepton with 
   $p_T \geq 8 \; \gevc$. The measurement is a comparison between the data 
   and a {\sc herwig}-based simulation of the semileptonic decay rate for
   the additional jets (away-jets), which have no lepton trigger requirement.
   We first use measured rates of lepton- and away-jets with SECVTX and JPB
   tags in order to determine the bottom and charmed content of the data; 
   we then tune the simulation to match the observed heavy-flavor content.
   Rates of SECVTX and JPB tags and the kinematics of these events are
   well modeled after tuning the parton-level cross sections predicted by
   {\sc herwig} within the experimental and theoretical uncertainties. The
   tuned parton-level prediction of {\sc herwig} indicates that, in order 
   to model the single $b$ production cross section measured at the Tevatron,
   any theoretical calculation should predict higher-order-term contributions
   which are approximately a factor of three larger than the LO contribution.

   We then measure the yields of soft ($p_T\geq 2\; \gevc$) leptons due to 
   heavy-flavor decays in the away-jets, and compare them to the prediction
   of the tuned simulation. The latter depends on the bottom and charmed 
   semileptonic decay rates and on the soft lepton reconstruction efficiency.
   To calibrate the predictions of the simulation, we perform the same
   analysis on samples of generic jets with $20$, $50$, and $100\; \gev$ 
   $E_T$ thresholds; these samples have also been previously used to
   calibrate the simulation of heavy flavor background to pair production
   of top quarks~\cite{cdf-tsig}.

   Finally, with these calibrations we find that away-jets have a $30 - 50$\%
   excess of soft lepton tags as compared with the simulation, corresponding
   to $2.5-3.5\; \sigma$, depending on the selection of the away-jets; the
   selections include (a) all away-jets, (b) a subset with SECVTX tags,
   and (c) another subset with JPB tags (the three results are highly
   correlated and should not be combined). The size of this excess is
   consistent with the differences between the NLO prediction and the  
   $b\bar{b}$ cross section measurements at the Tevatron that are based upon
   the detection of one and two leptons from $b$-quark decays. A possible 
   interpretation of this excess, the one that motivated this study, is the
   pair production of light scalar quarks with a 100\% semileptonic branching
   ratio. Due to the $p_T \geq 8\; \gevc$ lepton-trigger requirement, we 
   expected such a signature to be enhanced in this sample as compared with
   generic-jet data. However, alternative explanations for the excess are 
   not excluded by this study, the interpretation of which requires
   independent confirmations.
  \section{Acknowledgments}
   We thank the Fermilab staff and the technical staff of the participating
   Institutions for their contributions. This work was supported by the
   U.S.~Department of Energy and National Science Foundation; the Istituto
   Nazionale di Fisica Nucleare; the Ministry of Education, Culture, Sports,
   Science and Technology of Japan; the National Science Council of the
   Republic of China; the Swiss National Science Foundation; the 
   A.P.~Sloan Foundation; the Bundesministerium f\"{u}r Bildung und 
   Forschung; the Korea Science and Engineering Foundation (KoSEF); 
   the Korea Research Foundation; and the Comision Interministerial de 
   Ciencia y Tecnologia, Spain. We benefitted from many useful discussions
   with M.~Seymour and S.~Moretti.
 \input{bibliography_mod.tex}

 \end{document}

%% file: prd_definitions.tex
\def\deg{^\circ}

\def\W{{\em W\/ }}
\def\Z0{${\em Z^0\/}$}

\def\r#1 {$^{#1}$}

\newcommand{\gev}  { {\rm GeV}}

\newcommand{\gevc} { {\rm GeV/c}}
\newcommand{\gevcc}{ {\rm GeV/c^2}}

\def\gepsfcentered#1{
  \def\testit{#1}
  \def\lbracket{[}
  \ifx\testit\lbracket
    \let\dofilecmd=\gepsfwithopt
  \else
    \let\dofilecmd=\gepsfnoopt
  \fi
  \dofilecmd}

\def\gepsfnoopt#1{
  \begin{center}
  \leavevmode
  \epsffile{#1}
  \end{center}}

\def\gepsfwithopt#1 #2 #3 #4]#5{
  \begin{center}
  \leavevmode
  \gepsfmaxx=0.94\textwidth
  \epsffile[#1 #2 #3 #4]{#5}
  \end{center}}

\newdimen\gepsfmaxx
\def\epsfsize#1#2{
  \ifnum \epsfxsize=0
    \ifnum \epsfysize=0
      \ifnum #1 > \gepsfmaxx
        \gepsfmaxx
      \else
        #1
      \fi
    \else
      \epsfxsize
    \fi
  \else
    \epsfxsize
  \fi
}

%% file: auth_luc.tex
\font\eightit=cmti8
\def\r#1{\ignorespaces $^{#1}$}
\hfilneg
\begin{sloppypar}
\noindent
D.~Acosta,\r {11} D.~Ambrose,\r {31} K.~Anikeev,\r {23} J.~Antos,\r 1
G.~Apollinari,\r {10} T.~Arisawa,\r {42} A.~Artikov,\r 8
F.~Azfar,\r {29} P.~Azzi-Bacchetta,\r {30} N.~Bacchetta,\r {30}
% ...
\end{sloppypar}
\vskip .026in
\begin{center}
(CDF Collaboration)
\end{center}

\vskip .026in
\begin{center}
\r 1    {\eightit Institute of Physics, Academia Sinica, Taipei, Taiwan 11529, 
         Republic of China} \\
\r 2    {\eightit Istituto Nazionale di Fisica Nucleare, University of Bologna,
         I-40127 Bologna, Italy} \\
\r 3    {\eightit Brandeis University, Waltham, Massachusetts 02254} \\
\r 4    {\eightit University of California at Davis, Davis, California 95616} \\
\r 5    {\eightit University of California at Los Angeles, Los 
         Angeles, California  90024} \\ 
\r 6    {\eightit Instituto de Fisica de Cantabria, CSIC-University of 
         Cantabria, 39005 Santander, Spain} \\
\r 7    {\eightit Carnegie Mellon University, Pittsburgh, Pennsylvania 15213}\\
\r 8    {\eightit Joint Institute for Nuclear Research, RU-141980 
         Dubna, Russia} \\
\r 9    {\eightit Duke University, Durham, North Carolina  27708} \\
\r {10} {\eightit Fermi National Accelerator Laboratory, Batavia, Illinois 
         60510} \\
\r {11} {\eightit University of Florida, Gainesville, Florida  32611} \\
\r {12} {\eightit Laboratori Nazionali di Frascati, Istituto Nazionale di 
         Fisica Nucleare, I-00044 Frascati, Italy} \\
\r {13} {\eightit University of Geneva, CH-1211 Geneva 4, Switzerland} \\
\r {14} {\eightit Glasgow University, Glasgow G12 8QQ, United Kingdom}\\
\r {15} {\eightit Harvard University, Cambridge, Massachusetts 02138} \\
\r {16} {\eightit Hiroshima University, Higashi-Hiroshima 724, Japan} \\
\r {17} {\eightit University of Illinois, Urbana, Illinois 61801} \\
\r {18} {\eightit The Johns Hopkins University, Baltimore, Maryland 21218} \\
\r {19} {\eightit Institut f\"{u}r Experimentelle Kernphysik, 
         Universit\"{a}t Karlsruhe, 76128 Karlsruhe, Germany} \\
\r {20} {\eightit Center for High Energy Physics: Kyungpook National
         University, Taegu 702-701; Seoul National University, Seoul 151-742;
         and SungKyunKwan University, Suwon 440-746; Korea} \\
\r {21} {\eightit Ernest Orlando Lawrence Berkeley National Laboratory, 
         Berkeley, California 94720} \\
\r {22} {\eightit University College London, London WC1E 6BT, United Kingdom}\\
\r {23} {\eightit Massachusetts Institute of Technology, Cambridge,
         Massachusetts  02139} \\   
\r {24} {\eightit University of Michigan, Ann Arbor, Michigan 48109} \\
\r {25} {\eightit Institution for Theoretical and Experimental Physics, ITEP,
         Moscow 117259, Russia} \\
\r {26} {\eightit University of New Mexico, Albuquerque, New Mexico 87131} \\
\r {27} {\eightit Northwestern University, Evanston, Illinois  60208} \\
\r {28} {\eightit Osaka City University, Osaka 588, Japan} \\
\r {29} {\eightit University of Oxford, Oxford OX1 3RH, United Kingdom} \\
\r {30} {\eightit Universita di Padova, Istituto Nazionale di Fisica 
         Nucleare, Sezione di Padova, I-35131 Padova, Italy} \\
\r {31} {\eightit University of Pennsylvania, Philadelphia, 
         Pennsylvania 19104} \\   
\r {32} {\eightit Istituto Nazionale di Fisica Nucleare, University and Scuola
         Normale Superiore of Pisa, I-56100 Pisa, Italy} \\
\r {33} {\eightit Purdue University, West Lafayette, Indiana 47907} \\
\r {34} {\eightit University of Rochester, Rochester, New York 14627} \\
\r {35} {\eightit Rockefeller University, New York, New York 10021} \\
\r {36} {\eightit Instituto Nazionale de Fisica Nucleare, Sezione di Roma,
University di Roma I, ``La Sapienza," I-00185 Roma, Italy}\\
\r {37} {\eightit Texas A\&M University, College Station, Texas 77843} \\
\r {38} {\eightit Texas Tech University, Lubbock, Texas 79409} \\
\r {39} {\eightit Istituto Nazionale di Fisica Nucleare, University
         of Trieste/Udine, Italy} \\
\r {40} {\eightit University of Tsukuba, Tsukuba, Ibaraki 305, Japan} \\
\r {41} {\eightit Tufts University, Medford, Massachusetts 02155} \\
\r {42} {\eightit Waseda University, Tokyo 169, Japan} \\
\r {43} {\eightit University of Wisconsin, Madison, Wisconsin 53706} \\
\end{center}

%% file: tab_strat.tex
 \begin{table}
 \caption{Comparison between $\sigma= BR_b \times \sigma_{b\bar{b}}+
 	  BR_c \times \sigma_{c\bar{c}} + BR_{\tilde{b}} \times 
          \sigma_{\tilde{b}\tilde{b}^*}$, the total heavy-flavor production 
          cross section ($b$, $c$, and $\tilde{b}$) contributing to different
          hypothetical samples, and $\sigma^{norm}= BR_b \times
          \sigma^{norm}_{b\bar{b}} + BR_c \times \sigma_{c\bar{c}}$, the total
          heavy-flavor cross section determined with a conventional-QCD
          simulation under the hypothesis that scalar quarks have the same
          lifetime of $b$ quarks ($\sigma^{norm}_{b\bar{b}} = 
          \sigma_{b\bar{b}} + \sigma_{\tilde{b}\tilde{b}^*}$). In samples
          containing leptons, each cross section is also multiplied by the
          appropriate semileptonic branching ratio $BR$.} 
 \begin{center}
 \def\arraystretch{1.2}
 \begin{tabular}{lccc}
    {\scriptsize Sample}  & \multicolumn{1}{c}{\scriptsize $\sigma $ (nb)} 
  & \multicolumn{1}{c}{\scriptsize $\sigma^{norm}$ (nb)}  
                      & {\scriptsize $\sigma/\sigma^{norm}$ } \\ 
    {\scriptsize A $=$ generic jets}  & {\scriptsize $869 = 298 + 487 + 84$} 
  & {\scriptsize $869 = 382 + 487$}   & {\scriptsize 1.0 }     \\ 
    {\scriptsize B $=$ A with one lepton}  
  & {\scriptsize $ 296 = 0.37 \times 298 + 0.21\times 487 + 1.0\times 84$}
  & {\scriptsize $ 244 = 0.37 \times 382 + 0.21\times 487$}  
  & {\scriptsize 1.2}   \\ 
    {\scriptsize C $=$ A with two leptons} 
  & {\scriptsize $ 146 = 0.37^2 \times 298 + 0.21^2\times 487 + 1.0\times 84$}
  & {\scriptsize $ 74  = 0.37^2\times 382 + 0.21^2\times 487$} 
  & {\scriptsize 2.0} \\ 
    {\scriptsize D $=$ B renormalized} & {\scriptsize $296 = 110 + 102 + 84$}
  & {\scriptsize $ 296 = 194 + 102$}   & {\scriptsize 1.0} \\
    {\scriptsize E $=$ D with one lepton}  
  & {\scriptsize $ 146 = 0.37\times 110 + 0.21\times 102 + 1.0\times 84$} 
  & {\scriptsize $ 93  = 0.37\times 194 + 0.21\times 102$} 
  & {\scriptsize 1.5} \\       
  \end{tabular}
  \label{tab:tab_strat} 
  \end{center}
  \end{table}

%% file: datatag_1.tex
 \begin{table}
 \caption{Number of tags due to heavy flavors in the inclusive
          lepton data (raw counts/removed mistags are indicated in
          parenthesis). $P_{GQCD}$ is the probability of tagging away-jets
          recoiling against lepton-jets without heavy flavor.}
 \begin{center}
 \begin{tabular}{lcccccc}
 &\multicolumn{3}{c}{\bf Electron data}  &\multicolumn{3}{c}{\bf Muon data}\\ 
 Tag type          &                     &               & $P_{GQCD}$       
                   &                     &               & $P_{GQCD}$ \\
 $N_{l-jet}$       & $68544$             &               &                 
                   & $14966$             &               &            \\
 $N_{a-jet}$       & $73335$             &               &                
                   & $16460$             &               &            \\
 $T_{l-jet}^{SEC}$ & $10115.3 \pm 101.7$ & (10221/105.7) &   
                   & $3657.3  \pm 60.8$  &  (3689/31.7)~ &            \\
 $T_{l-jet}^{JPB}$ & $11165.4 \pm 115.8$ & (11591/425.6) &  
                   & $4068.6  \pm 66.2$  &  (4204/135.4) &            \\
 $T_{a-jet}^{SEC}$ & $~4353.3 \pm 68.5~$ & ~(4494/140.7) &  $1.56\%$  
                   & $1054.6  \pm 33.3$  &  (1094/39.4)~ &  $1.67\%$  \\
 $T_{a-jet}^{JPB}$ & $~5018.9 \pm 98.9~$ & ~(5661/642.1) &  $2.45\%$  
                   & $1265.2  \pm 41.1$  &  (1427/161.8) &  $2.63\%$  \\
 $DT^{SEC}$        & $~1375.2 \pm 37.6~$ & ~(1405/29.8)~ &  
                   & $~452.6  \pm 21.6$  &  ~(465/12.4)~ &            \\
 $DT^{JPB}$        & $~1627.8 \pm 43.7~$ & ~(1754/126.2) &  
                   & $~546.4  \pm 25.1$  &  ~(600/53.6)~ & \\
 \end{tabular}
 \end{center}
 \label{tab:datatag_1}
 \end{table}

%% file: tab_simtag_1.tex
 \begin{table}
 \caption{Number of jets before and after tagging in the inclusive lepton
          simulation (dir, f.exc and gsp indicate the direct production,
          flavor excitation and gluon splitting contributions). The row
          indicated as ``h.f./light" lists the rates of away-jets with and
          without heavy flavors and highlights the properties of different
          production mechanisms. Data-to-simulation scale factors for the
          various tagging algorithms are not yet applied.}
 \begin{center}
 \begin{tabular}{lcccccc}
          & \multicolumn{5}{c}{\bf Electron simulation} & \\
 Tag type           &$b$-dir  &$c$-dir&$b$-f.exc &$c$-f.exc&$b$-gsp &$c$-gsp \\
 $HF_{l-jet}$       &   5671  &  947  &   10779  &  2786   &  5263  & 1690   \\
 $HF_{a-jet}$       &   5848  &  977  &   11280  &  2913   &  6025  & 1877   \\
 h.f./light         & 5407/441& 899/78& 1605/9675&367/2546 &707/5318&145/1732\\
 $HFT_{l-jet}^{SEC}$&   1867  &  ~52  &   ~3624  &  ~194   &  1732  & ~147   \\
 $HFT_{l-jet}^{JPB}$&   2392  &  163  &   ~4531  &  ~602   &  2106  & ~356   \\
 $HFT_{a-jet}^{SEC}$&   2093  &  ~91  &   ~~480  &  ~~68   &  ~222  & ~~15   \\
 $HFT_{a-jet}^{JPB}$&   2622  &  203  &   ~~584  &  ~136   &  ~276  & ~~58   \\
 $HFDT^{SEC}$       &   ~678  &  ~~5  &   ~~157  &  ~~~4   &  ~~78  & ~~~1   \\
 $HFDT^{JPB}$       &   1083  &  ~43  &   ~~303  &  ~~25   &  ~168  & ~~18   \\
          & \multicolumn{5}{c}{\bf Muon simulation} & \\
 Tag type           &$b$-dir  &$c$-dir&$b$-f.exc &$c$-f.exc&$b$-gsp &$c$-gsp \\
 $HF_{l-jet}$       &   1285  &  298  &   ~2539  &  ~942   &  1455  & ~747   \\
 $HF_{a-jet}$       &   1358  &  313  &   ~2705  &  ~994   &  1708  & ~816   \\
 h.f./light         & 1206/152& 278/35& 422/2283 & 124/870 &171/1537& 48/768 \\
 $HFT_{l-jet}^{SEC}$&   ~569  &  ~34  &   ~1131  &  ~~83   &  ~652  & ~~92   \\
 $HFT_{l-jet}^{JPB}$&   ~707  &  ~77  &   ~1386  &  ~229   &  ~830  & ~202   \\
 $HFT_{a-jet}^{SEC}$&   ~498  &  ~29  &   ~~132  &  ~~13   &  ~~54  & ~~11   \\
 $HFT_{a-jet}^{JPB}$&   ~627  &  ~62  &   ~~173  &  ~~34   &  ~~60  & ~~21   \\
 $HFDT^{SEC}$       &   ~218  &  ~~3  &   ~~~59  &  ~~~2   &  ~~20  & ~~~1   \\
 $HFDT^{JPB}$       &   ~347  &  ~12  &   ~~105  &  ~~~7   &  ~~50  & ~~~6   \\
 \end{tabular}
 \end{center}
 \label{tab:simtag_1}
 \end{table}

%% file: tab_norm_1.tex
 \begin{table}
 \caption[]{Result of the fit of the {\sc herwig} simulation to the data. 
	    The fit is described in the text and yields $\chi^{2}/{\rm DOF}
            = 4.6/9$. The rescaling factors for the gluon splitting 
            contributions predicted by the {\sc herwig} parton-shower
            Monte Carlo are of the same size as those measured by the SLC 
            and LEP experiments~\cite{balles}, and are consistent with the
            estimated theoretical uncertainty~\cite{seymour}. }
 \begin{center}
 \begin{tabular}{llc}
  SECVTX scale factor      & $SF_{b}$   & $0.97 \pm 0.03$\\
  SECVTX scale factor      & $SF_{c}$   & $0.94 \pm 0.22$\\
  JPB scale factor         & $SF_{JPB}$ & $1.01 \pm 0.02$\\
  $e$  norm.               & $K_{e}$    & $1.02 \pm 0.05$\\
  $\mu$  norm.             & $K_{\mu}$  & $1.08 \pm 0.06$\\
  $c$ dir. prod.           & $c$        & $1.01 \pm 0.10$\\
  $b$ flav. exc.           & $bf$       & $1.02 \pm 0.12$\\
  $c$ flav. exc.           & $cf$       & $1.10 \pm 0.29$\\
  $g \rightarrow b\bar{b}$ & $bg$       & $1.40 \pm 0.18$\\ 
  $g \rightarrow c\bar{c}$ & $cg$       & $1.40 \pm 0.34$\\
 \end{tabular}
 \end{center}
 \label{tab:norm_1}
 \end{table}

%% file: tab_norm_1a.tex
 \begin{table}
 \caption{Parameter correlation coefficients.} 
 \begin{center}
 \begin{tabular}{cccccccccc}
           &$SF_{c}$&$SF_{JPB}$&$K_{e}$&$c$&$bf$&$cf$&$bg$ &$cg$&$K_{\mu}$ \\
 $SF_{b}$  & $-0.073$ & $~0.718$ & $-0.747$ & $~0.054$ & $~0.346$ & $~0.297$
           & $-0.062$ & $~0.066$ & $-0.715$\\
 $SF_{c}$  &          & $~0.358$ & $-0.238$ & $-0.002$ & $~0.038$ & $~0.147$
           & $-0.071$ & $~0.086$ & $-0.306$\\
 $SF_{JPB}$&          &          & $-0.810$ & $~0.010$ & $~0.363$ & $~0.127$
           & $-0.009$ & $-0.049$ & $-0.802$\\
 $K_{e}$   &          &          &          & $-0.092$ & $-0.641$ & $-0.302$
           & $~0.071$ & $~0.077$ & $~0.933$\\
 $c$       &          &          &          &          & $~0.053$ & $~0.020$
           & $~0.008$ & $~0.002$ & $-0.098$\\
 $bf$      &          &          &          &          &          & $~0.245$
           & $-0.680$ & $-0.199$ & $-0.526$\\
 $cf$      &          &          &          &          &          &   
           & $-0.321$ & $-0.164$ & $-0.274$\\
 $bg$      &          &          &          &          &          &   
           &          & $-0.029$ & $-0.019$\\
 $cg$      &          &          &          &          &          &   
           &          &          & $-0.018$\\
 \end{tabular}
 \end{center}
 \label{tab:norm_1a} 
 \end{table}

%% file: tab_norm_2.tex
 \begin{table}
 \caption{Rates of tags due to heavy flavor in the data and in the fitted
          {\sc herwig} simulation. The heavy flavor purity of the lepton-jets
          in the data returned by the best fit is $F_{hf} =(45.3 \pm 1.9)$\%
          in the electron sample and $F_{hf} = (59.7 \pm 3.6)$\% in the muon
          sample. The contribution of a-jets recoiling against l-jets without
          heavy flavor has been subtracted; the 10\% uncertainty of this 
          contribution is included in the errors.}
 \begin{center}
 \begin{tabular}{lcccc}
     & \multicolumn{2}{c}{\bf Electrons} &\multicolumn{2}{c}{\bf Muons}\\
  Tag type             & Data             & Simulation & Data & Simulation\\
 $HFT_{l-jet}^{SEC}$   & $10115.3 \pm 101.7$ & $10156.8 \pm  159.3$  
                       &  $3657.3 \pm  60.8$ &  $3636.7 \pm   95.8$\\
 $HFT_{l-jet}^{JPB}$   & $11165.4 \pm 115.8$ & $11139.8 \pm  159.7$
                       &  $4068.6 \pm  66.2$ &  $4059.7 \pm   95.8$\\
 $HFT_{a-jet}^{SEC}$   & $~3729.0 \pm 92.8~$ & $~3691.5 \pm  109.7$   
                       &  $~943.8 \pm  35.2$ &  $~967.4 \pm   43.2$\\
 $HFT_{a-jet}^{JPB}$   & $~4035.8 \pm 139.7$ & $~3984.0 \pm  111.0$   
                       &  $1090.8 \pm  44.9$ &  $1059.3 \pm   42.8$\\
 $HFDT^{SEC}$          & $~1375.2 \pm 37.6~$ & $~1380.8 \pm  59.4~$     
                       &  $~452.6 \pm  21.6$ &  $~474.3 \pm   31.1$\\
 $HFDT^{JPB}$          & $~1627.8 \pm 43.7~$ & $~1644.0 \pm  57.1~$
                       &  $~546.4 \pm  25.1$ &  $~556.6 \pm   28.7$\\
 \end{tabular}
 \end{center}
 \label{tab:tab_norm_2}
 \end{table}

%% file: tab_ratslt_1.tex
 \begin{table}
 \caption{Number of away-jets with SLT tags due to heavy flavors in the 
          inclusive lepton sample. Raw counts and removed mistags are listed
          in parentheses. When appropriate, mistags include fake 
          SECVTX (JPB) contributions. $P_{GQCD}$ is the probability of 
          finding a tag due to heavy flavor in away-jets recoiling against 
          a lepton-jet without heavy flavor.}
 \begin{center}
 \begin{tabular}{lcccccc}
   &\multicolumn{3}{c}{\bf Electron data} &\multicolumn{3}{c}{\bf Muon data}\\ 
 Tag type & & & $P_{GQCD}$& & & $P_{GQCD}$ \\
 $T_{a-jet}^{SLT}$            & $1063.8 \pm 113.0$ &($2097/1033.2$)& $0.49\%$
                              &  $308.6 \pm  34.7$ & ($562/253.4$) & $0.54\%$\\
 $T_{a-jet}^{SLT\cdot SEC}$   & $~356.3 \pm 22.8~$ &~($444/87.7$)~~& $0.08\%$
                              &  $~69.3 \pm  9.9~$ & ~($92/22.7$)~ & $0.09\%$\\
 $T_{a-jet}^{SLT\cdot JPB}$   & $~401.3 \pm 25.3~$ &~($513/111.7$)~& $0.13\%$
                              &  $112.3 \pm  12.3$ & ($143/30.7$)~ & $0.14\%$\\
 \end{tabular}
 \end{center}
 \label{tab:tab_rateslt_1}
 \end{table}

%% file: tab_ratslt_2.tex
 \begin{table}
 \caption{Rates of away-jets with SLT tag due to heavy flavors in the 
	  inclusive lepton simulation. The data-to-simulation scale factor
          for the supertag efficiency is not yet applied.}
 \begin{center}
 \begin{tabular}{lcccccc}
         & \multicolumn{5}{c}{\bf Electron simulation} & \\
   Tag type                   &$b$-dir  &$c$-dir &$b$-f.exc &$c$-f.exc &$b$-gsp
   & $c$-gsp \\
 $HFT_{a-jet}^{SLT}$          &   362   &   26   &   93     &  30      &  41
   &  9\\
 $HFT_{a-jet}^{SLT\cdot SEC}$ &   159   &    1   &   47     &   2      &  18
   &  0\\
 $HFT_{a-jet}^{SLT\cdot JPB}$ &   200   &    7   &   53     &   6      &  21
   &  2\\
         & \multicolumn{5}{c}{\bf Muon simulation} & \\
 $HFT_{a-jet}^{SLT}$          &    82   &   10   &   21     &   5      &   9
   &  5\\
 $HFT_{a-jet}^{SLT\cdot SEC}$ &    33   &    2   &    9     &   0      &   4
   &  0\\
 $HFT_{a-jet}^{SLT\cdot JPB}$ &    44   &    3   &   13     &   3      &   5
   &  2\\
 \end{tabular}
 \end{center}
 \label{tab:tab_rateslt_2}
 \end{table}

%% file: tab_ratslt_5n.tex
 \begin{table}
 \caption{Number of a-jets with an SLT tag due to heavy flavor decay. The 
          contribution of a-jets recoiling against l-jets without heavy 
          flavor has been subtracted (see text).}
 \begin{center}
 \def\arraystretch{1.2}
 \begin{tabular}{lcccc}
     & \multicolumn{2}{c}{\bf Electrons} &\multicolumn{2}{c}{\bf Muons}\\
   Tag type                   & Data                    & Simulation     
                              & Data                    & Simulation\\
 $HFT_{a-jet}^{SLT}$          & $865.1 \pm 114.8$       &  $597.6 \pm 69.3$ 
                              & $272.7 \pm  34.9$       &  $149.3 \pm 21.0$\\
 $HFT_{a-jet}^{SLT\cdot SEC}$ & $322.6 \pm 23.3~$       &  $242.4 \pm 22.5$
                              & $~63.3 \pm  9.9~$       &  $~53.8 \pm 8.7~$\\
 $HFT_{a-jet}^{SLT\cdot JPB}$ & $350.2 \pm 26.3~$       &  $251.5 \pm 21.7$
                              & $103.2 \pm  12.4$       &  $~65.0 \pm 8.9~$\\
 \end{tabular}
 \end{center}
 \label{tab:tab_rateslt_5}
 \end{table}

%% file: tab_ratslt_6n.tex
 {\squeezetable
 \begin{table}
 \caption{Tagging rates in the normalized simulation listed by production
          mechanisms.}
 \begin{center}
 \def\arraystretch{1.0}
 \begin{tabular}{lcccccc}
      & & & \multicolumn{2}{c}{\bf Electron simulation} & & \\
    Tag type     & $b$-dir &$c$-dir &$b$-f.exc &$c$-f.exc &$b$-gsp  &$c$-gsp\\ 
 $HF_{l-jet}$       &$5781.0\pm320.8$ & $~973.2\pm 109.8$ & $11247.8\pm1073.9$ 
                    &$3115.7\pm790.1$ & $7504.6\pm1081.6$ & $2411.0\pm593.8$\\
 $HF_{a-jet}$       &$5961.4\pm330.6$ & $1004.0\pm 113.2$ & $11770.6\pm1123.6$
                    &$3257.7\pm826.0$ & $8591.1\pm1237.4$ & $2677.8\pm659.2$\\
 $HFT_{l-jet}^{SEC}$&$2267.5\pm101.6$ & $~~49.1\pm19.4~$  & $~4505.5\pm451.7~$
                    &$~199.5\pm81.7~$ & $2942.4\pm408.7~$ & $~192.8\pm87.8~$\\
 $HFT_{l-jet}^{JPB}$&$2358.3\pm99.0~$ & $~162.0\pm20.7~$  & $~4572.8\pm454.2~$
                    &$~651.1\pm167.3$ & $2904.3\pm404.4~$ & $~491.2\pm122.2$\\
 $HFT_{a-jet}^{SEC}$&$2542.0\pm112.3$ & $~~86.0\pm33.1~$  & $~~596.8\pm65.0~~$ 
                    &$~~69.9\pm29.4~$ & $~377.1\pm57.5~~$ & $~~19.7\pm10.2~$\\
 $HFT_{a-jet}^{JPB}$&$2585.1\pm107.3$ & $~201.8\pm24.8~$  & $~~589.4\pm62.8~~$ 
                    &$~147.1\pm39.4~$ & $~380.6\pm57.1~~$ & $~~80.0\pm22.1~$\\
 $HFDT^{SEC}$       &$~981.1\pm52.5~$ & $~~~4.3\pm3.6~~$  & $~~232.5\pm31.4~~$ 
                    &$~~~3.8\pm3.3~~$ & $~157.9\pm27.8~~$ & $~~~1.2\pm1.5~~$\\
 $HFDT^{JPB}$       &$1032.7\pm45.8~$ & $~~41.3\pm7.5~~$  & $~~295.7\pm36.0~~$
                    &$~~26.2\pm8.5~~$ & $~224.1\pm35.0~~$ & $~~24.0\pm8.1~~$\\
 $HFT_{a-jet}^{SLT}$&$~369.0\pm46.2~$ & $~~26.7\pm6.6~~$  & $~~~97.0\pm16.7~~$
                    &$~~33.6\pm11.0~$ & $~~58.5\pm13.7~~$ & $~~12.8\pm5.5~~$\\
 $HFT_{a-jet}^{SLT\cdot SEC}$ 
                    &$~164.1\pm17.4~$ & $~~~0.8\pm0.9~~$  & $~~~49.7\pm9.2~~~$
                    &$~~~1.7\pm1.4~~$ & $~~26.0\pm7.2~~~$ & $~~~~~~0~~~~~$\\
 $HFT_{a-jet}^{SLT\cdot JPB}$
                    &$~167.6\pm16.6~$ & $~~~5.9\pm2.3~~$  & $~~~45.5\pm8.1~~~$
                    &$~~~5.5\pm2.7~~$ & $~~24.6\pm6.5~~~$ & $~~~2.3\pm1.8~~$\\
    & & & \multicolumn{2}{c}{\bf Muon simulation} & &\\
    Tag type    & $b$-dir &$c$-dir &$b$-f.exc &$c$-f.exc &$b$-gsp  &$c$-gsp\\ 
 $HF_{l-jet}$       &$1383.7\pm84.4~$ & $~323.5\pm39.6~$  & $2798.6\pm292.4~$
                    &$1112.8\pm285.4$ & $2191.5\pm310.5~$ & $1125.7\pm284.9$\\
 $HF_{a-jet}$       &$1462.3\pm88.7~$ & $~339.8\pm41.4~$  & $2981.5\pm311.2~$
                    &$1174.2\pm301.0$ & $2572.5\pm363.5~$ & $1229.7\pm310.9$\\
 $HFT_{l-jet}^{SEC}$&$~730.0\pm42.3~$ & $~~33.9\pm14.0~$  & $1485.2\pm164.3~$
                    &$~~90.1\pm38.0~$ & $1170.0\pm161.8~$ & $~127.5\pm59.3~$\\
 $HFT_{l-jet}^{JPB}$&$~736.3\pm39.0~$ & $~~80.8\pm12.3~$  & $1477.5\pm160.9~$
                    &$~261.6\pm69.0~$ & $1209.1\pm166.3~$ & $~294.4\pm76.0~$\\
 $HFT_{a-jet}^{SEC}$&$~638.9\pm38.4~$ & $~~28.9\pm12.1~$  & $~173.3\pm23.8~~$
                    &$~~14.1\pm7.0~~$ & $~~96.9\pm18.4~~$ & $~~15.2\pm8.3~~$\\
 $HFT_{a-jet}^{JPB}$&$~653.0\pm35.6~$ & $~~65.1\pm10.5~$  & $~184.4\pm24.0~~$
                    &$~~38.8\pm11.9~$ & $~~87.4\pm16.2~~$ & $~~30.6\pm10.1~$\\
 $HFDT^{SEC}$       &$~333.2\pm26.2~$ & $~~~2.8\pm2.5~~$  & $~~92.3\pm16.1~~$
                    &$~~~2.0\pm2.0~~$ & $~~42.8\pm11.1~~$ & $~~~1.3\pm1.6~~$\\
 $HFDT^{JPB}$       &$~349.5\pm22.0~$ & $~~12.2\pm3.7~~$  & $~108.3\pm16.3~~$
                    &$~~~7.7\pm3.5~~$ & $~~70.4\pm13.6~~$ & $~~~8.5\pm4.0~~$\\
 $HFT_{a-jet}^{SLT}$&$~~88.3\pm14.0~$ & $~~10.9\pm3.8~~$  & $~~23.1\pm6.0~~~$
                    &$~~~5.9\pm3.1~~$ & $~~13.6\pm5.1~~~$ & $~~~7.5\pm3.9~~$\\
 $HFT_{a-jet}^{SLT\cdot SEC}$
                    &$~~36.0\pm6.8~~$ & $~~~1.7\pm1.4~~$  & $~~10.0\pm3.6~~~$
                    &$~~~~~~0~~~~~$   & $~~~6.1\pm3.2~~~$ & $~~~~~~0~~~~~$\\
 $HFT_{a-jet}^{SLT\cdot JPB}$
                    &$~~38.9\pm6.5~~$ & $~~~2.7\pm1.6~~$  & $~~11.8\pm3.6~~~$
                    &$~~~2.9\pm1.8~~$ & $~~~6.2\pm2.9~~~$ & $~~~2.5\pm1.9~~$\\
 \end{tabular}
 \end{center}
 \label{tab:tab_rateslt_6}
 \end{table}
 }

%% file: tab_ratslt_sum.tex
 \begin{table}
 \caption{Summary of the observed and predicted numbers of a-jets with SLT
          tags or supertags in the inclusive lepton sample. Mistags are the 
          expected fake-tag contributions in a-jets recoiling against l-jets
          with heavy flavor (h.f.). QCD are the predicted numbers of tags, 
          which include mistags, in a-jets recoiling l-jets without heavy
          flavor. $HFT_{a_jet}$ (data and h.f. simulation) are the numbers
          of tagged a-jets with heavy flavor recoiling against l-jets with
          heavy flavor; in the data, this contribution is obtained by 
          subtracting the second plus third rows of this table from the 
          first one.}
 \begin{center}
 \def\arraystretch{1.2}
 \begin{tabular}{lcccc}
   Tag type                     &      SLT      &  SLT+SECVTX  &  SLT+JPB   \\
 Observed			&     $2659$    &    $536$     &    $656$   \\
 Mistag                         & $~619 \pm62~$ & $~53 \pm 5~$ & $~69 \pm 7~$\\
 QCD				& $~902 \pm91~$ & $~97 \pm 10$ & $134 \pm 13$\\
 $HFT_{a-jet}$ (data)           & $1138 \pm140$ & $386 \pm 26$ & $453 \pm 29$\\
 $HFT_{a-jet}$ (h.f.simulation) & $~747 \pm75~$ & $296 \pm 26$ & $317 \pm 25$\\
 Excess                         & $~391 \pm159$ & $~90 \pm 37$ & $136 \pm 38$\\
 \end{tabular}
 \end{center}
 \label{tab:tab_rateslt_sum}
 \end{table}

%% file: tab_4_1.tex
 \begin{table}
 \begin{center}
 \def\arraystretch{0.9}
 \caption[]{Number of tags due to heavy flavors in three samples of generic
            jets~\cite{jet} and in their tuned simulation. The amount of
            mistags removed from the data is indicated in parenthesis; 
            errors include a 10\% uncertainty in the mistag evaluation.
            The yields of tags in the simulation have been corrected with 
            the appropriate scale factors (see Sec.~\ref{sec:ss-sample}).
            The error of the number of simulated SLT tags includes the 10\%
            uncertainty of the SLT tagging efficiency in the simulation;
            the simulation efficiency for finding supertags (SLT+ SECVTX and
            SLT+ JPB) has been empirically reduced by 15\% to reproduce
            generic-jet data with a 6\% accuracy.}
 \begin{tabular}{lcc}
         \multicolumn{3}{c}{ JET 20 (194,009 events) } \\
  Tag type   &    Data (removed fakes)      &    Simulation \\
 SECVTX      & $~~4058\pm 92~$ ~($616.0$)  &  $4052 \pm 143$  \\
 JPB         & $~~5542\pm 295$ ($2801.0$)  &  $5573 \pm 173$  \\
 SLT         & $~~1032\pm 402$ ($3962.0$)  &  $~826 \pm 122$  \\
 SLT+SECVTX  & $~219.8\pm 20~$ ~~($94.2$)  &  $~223 \pm 16~$  \\
 SLT+JPB     & $~287.3\pm 28~$ ~($166.7$)  &  $~280 \pm 19~$  \\
 \hline
         \multicolumn{3}{c}{ JET 50 (151,270 events) } \\
 Tag type    &    Data (removed fakes)      &    Simulation \\
 SECVTX      & $~~5176 \pm 158$ ($1360.0$)  &  $5314 \pm 142$ \\
 JPB         & $~~6833 \pm 482$ ($4700.0$)  &  $6740 \pm 171$ \\
 SLT         & $~~1167 \pm 530$ ($5241.0$)  &  $1116 \pm 111$ \\
 SLT+SECVTX  & $~~~347 \pm 29~$ ~($169.0$)  &  $~343 \pm 23~$ \\
 SLT+JPB     & $~427.5 \pm 42~$ ~($288.5$)  &  $~416 \pm 27~$ \\
 \hline
        \multicolumn{3}{c}{ JET 100 (129,434 events) } \\
 Tag type    &    Data (removed fakes)      & Simulation \\
 SECVTX      & $~~5455 \pm 239$ ($2227.0$)  & $5889 \pm 176$  \\
 JPB         & $~~6871 \pm 659$ ($6494.0$)  & $7263 \pm 202$  \\
 SLT         & $~~1116 \pm 642$ ($6367.0$)  & $1160 \pm 168$  \\
 SLT+SECVTX  & $~377.6 \pm ~36$ ~($243.4$)  & $~432 \pm 29~$  \\
 SLT+JPB     & $~451.8 \pm ~55$ ~($401.2$)  & $~478 \pm 32~$  \\
 \end{tabular}
 \label{tab:tab_4.1}
 \end{center}
 \end{table}

%% file: tab_4_1bis.tex
 \begin{table}[p]
 \begin{center}
 \caption[]{Number of SLT tags in all generic-jets listed in 
            Table~\ref{tab:tab_4.1} (sample~A) and in away-jets recoiling a 
            lepton-jet with heavy flavor (sample~D). Samples~B and~C are
            generic jets tagged with the SECVTX and JPB algorithms, 
            respectively. Before tagging with the SLT algorithm, the heavy
            flavor purity is 13\% for sample~A, 78\% for sample~B, 58\% for
            sample~C, and 26\% for the sample~D used in this study. The 
            prediction of the fake SLT rate is calculated with the same
            parametrized probability for all samples; the heavy flavor (h.f.)
            contributions are predicted with the same simulation.}
 \def\arraystretch{1.2}
 {\footnotesize
 \begin{tabular}{lccc}
  Sample                &Number of SLT tags& Predicted fakes  &Predicted h.f.\\
  A: JET 20+JET 50+JET 100         & 18885 & $15570 \pm 1557$ & $3102 \pm403$\\
  B: generic jets with SECVTX tags & ~1451 & $~~507 \pm 51~~$ & $~998 \pm60~$\\
  C: generic jets with JPB tags    & ~2023 & $~~856 \pm 86~~$ & $1174 \pm71~$\\
  D: away-jets                     & ~1757 & $~~619 \pm 62~~$ & $~747 \pm75~$\\
 \end{tabular}                              
 }
 \label{tab:tab_4.1bis}
 \end{center}
 \end{table}

%% file: tab_ratslt_7.tex
 \begin{table}
 \caption{Number of $J/\psi$ mesons identified in the data and in
          the fitted simulation.}
 \begin{center}
 \begin{tabular}{lcccc}
   & \multicolumn{2}{c}{\bf Electrons} &\multicolumn{2}{c}{\bf Muons}\\
        Tag type     &       Data       &    Simulation    &       Data 
 & Simulation\\
  $Dil_{\psi}$       & $176.0 \pm 14.4$ & $155.2 \pm 21.5$ &  $83.0 \pm 9.4$ 
 & $ 54.0 \pm 10.1$\\
  $Dil_{\psi}^{SEC}$ & $~57.8 \pm 8.8~$ & $~71.8 \pm 10.7$ &  $31.9 \pm 5.8$
 & $ 28.7 \pm 6.2~$\\
  $Dil_{\psi}^{JPB}$ & $~61.2 \pm 8.4~$ & $~68.9 \pm 9.4~$ &  $29.6 \pm 5.7$
 & $ 33.0 \pm 6.4~$\\
 \end{tabular}
 \end{center}
 \label{tab:tab_rateslt_7}
 \end{table}

%% file: bibliography_mod.tex

%% file: prd_ajets_v9.bbl
\begin{thebibliography}{99}
 \label{bibliography}
 \bibitem{suj}      D.~Acosta {\it et al.,} Phys.~Rev.~{\bf D65},
		    052007 (2002).
 \bibitem{nde}      P.~Dawson {\it et al.,} Nucl.~Phys.~{\bf B327}, 49
		    (1988).
 \bibitem{berger}   E.~L.~Berger {\it et al.,} Phys.~Rev.~Lett.~{\bf 86},
		    4231 (2001).
 \bibitem{braaten}  E.~Braaten {\it et al.,} Phys.~Rev.~{\bf D66},
		    034003 (2002).
 \bibitem{mnr}      M.~Mangano {\it et al.,} Nucl.~Phys.~{\bf B373}, 295 
	            (1992). The {\sc fortran} code is available from the 
		    authors, and was used to evaluate the cross sections for 
                    these particular topologies.
 \bibitem{derwent}  F.~Abe {\it et al.,} Phys.~Rev.~{\bf D53}, 1051 (1996).
 \bibitem{yuntae}   F.~Abe {\it et al.,} Phys.~Rev.~{\bf D55}, 2547 (1997).
 \bibitem{abbot}    B.~Abbot {\it et al.,} Phys.~Lett.~{\bf B487}, 264
	            (2000).
 \bibitem{frix}     S.~Frixione {\it et al.,} 
		    Adv.~Ser.~Direct.~High~Energy~Phys.~{\bf15}, 609
		    (1998).
 \bibitem{jung}     H.~Jung, Journ.~Phys.~{\bf G28}, 971 (2002). 
 \bibitem{mssm-can} H.~E.~Haber and G.~L.~Kane, Phys.~Rep.~{\bf 117C}, 
                    76 (1985).
 \bibitem{nappi}    C. Nappi, Phys.~Rev.~{\bf D25}, 84 (1982).
 \bibitem{carena}   M.~Carena {\it et al.,} Phys.~Rev.~Lett.~{\bf 86}, 
		    1963 (2001).
 \bibitem{cdf-tsig} T.~Affolder {\it et al.,} Phys.~Rev.~{\bf D64}, 
		    032002 (2001).
 \bibitem{herwig}   G.~ Marchesini and B.~R.~Webber, Nucl.~Phys.~{\bf B310}, 
		    461 (1988); G.~Marchesini {\it et al.,} 
                    Comput.~Phys.~Commun.~{\bf 67}, 465 (1992).
 \bibitem{jpb}      D.~Buskulic {\it et al.,} Phys.~Lett.~{\bf B313}, 
		    535 (1993).
 \bibitem{prosp}    W.~Beenakker {\it et al.,} Nucl.~Phys.~{\bf B492}, 
		    51 (1997);  hep-ph/9611232.
 \bibitem{mrsg}     A.~D.~Martin, R.~G.~Roberts and W.~J.~Stirling, 
		    Phys.~Lett.~{\bf B354}, 155 (1995).
 \bibitem{hf-sem}   Review of Particle Physics, D.~E.~Groom {\it et al.,} 
		    Eur.~Phys.~J.~{\bf C15}, 1 (2000).
 \bibitem{cdf-det}  F.~Abe {\it et al.,} Nucl.~Inst.~and~Methods~{\bf A271},
		    387 (1988).
 \bibitem{svx-det}  D.~Amidei {\it et al.,} 
                    Nucl.~Instrum.~Methods~Phys.~Res.~{\bf A350}, 73 (1994).
 \bibitem{mixing}   F.~Abe {\it et al.,} Phys.~Rev.~{\bf D60}, 051101 (1999).
 \bibitem{jet_clus} F.~Abe {\it et al.,} Phys.~Rev.~{\bf D45}, 1448 (1992).
 \bibitem{svx}      An SVX track is reconstructed in the central tracking 
		    chamber (CTC) with at least two associated SVX hits. 
                    It is refitted using the SVX hits and the CTC track
	            parameters and covariance matrix.
 \bibitem{cdf-evid} F.~Abe {\it et al.,} Phys.~Rev.~{\bf D50}, 2966 (1994).
 \bibitem{kestenb}  D.~Kestenbaum, Ph.D. Thesis (unpublished), Harvard 
		    University (1996).
 \bibitem{cleo}     Version 9\_1 of the CLEO simulation; P.~Avery, K.~Read,
	            G.~Trahern, Cornell Internal Note CSN-212, March 25, 1985
                    (unpublished).
 \bibitem{field}    R.~D.~Field, Phys.~Rev.~{\bf D65}, 094006 (2002).
 \bibitem{balles}   A.~Ballestrero {\it et al.,} hep-ph/006259; 
                    M.~Mangano, hep-ph/9911256.
 \bibitem{seymour}  M.~Seymour, Nucl.~Phys.~{\bf B426}, 163 (1995); 
                    M.~Mangano, Nucl.~Phys.~{\bf B405}, 536 (1993).
 \bibitem{jet}      JET NN data are collected with a Level 2 trigger which 
		    requires a calorimetry cluster with transverse energy
                    larger than NN GeV.
 \bibitem{dkapi}    F.~Abe {\it et al.,} Phys.~Rev.~{\bf D58}, 092002 (1998);
                    Phys.~Rev.~Lett.~{\bf 76}, 4462 (1996).
 \bibitem{drelly}   F.~Abe {\it et al.,} Phys.~Rev.~{\bf D59}, 052002 (1999);
	            Phys.~Rev.~{\bf D49}, 1 (1994).
 \bibitem{psi-sec}  F.~Abe {\it et al.,} Phys.~Rev.~Lett.~{\bf 79}, 573 
		    (1997); Phys.~Rev.~Lett.~{\bf 79}, 578 (1997).
 \bibitem{wenzel}   F.~Abe {\it et al.,} Phys.~Rev.~{\bf D57}, 5382 (1998);
	            Phys.~Rev.~{\bf D55}, 1142 (1997).
 \end{thebibliography}
